\documentclass[aps,prb,twocolumn,floatfix,showpacs]{revtex4}
\usepackage{graphicx}
\usepackage{natbib}
\usepackage{epstopdf}
\usepackage{amsmath}
\usepackage{amssymb}
\usepackage{mathbbol}

\usepackage{xcolor}

\graphicspath{{./finalfigs/}}

\newcommand{\beq}{\begin{equation}}
\newcommand{\eeq}{\end{equation}}
\newcommand{\beqy}{\begin{eqnarray}}
\newcommand{\eeqy}{\end{eqnarray}}
\newcommand{\beqs}{\begin{equation*}}
\newcommand{\eeqs}{\end{equation*}}
\newcommand{\bpm}{\begin{pmatrix}}
\newcommand{\epm}{\end{pmatrix}}
\newcommand{\C}{\mathbb{C}}
\newcommand{\R}{\mathbb{R}}

\begin{document}

\title{Keldysh field theory for nonequilibrium condensation in a parametrically
 pumped polariton system}
\author{K. Dunnett}
\author{M. H. Szyma\'nska}
\email{m.szymanska@ucl.ac.uk}
\affiliation{Department of Physics and Astronomy, University College London,
Gower Street, London, WC1E 6BT, United Kingdom}
\date{\today}


\begin{abstract}
We develop a quantum field theory for parametrically pumped polaritons using
Keldysh Green's function techniques. By considering the mean-field and Gaussian
fluctuations, we find that the low energy physics of the highly non-equilibrium
phase transition to the optical parametric oscillator regime is in many ways
similar to equilibrium condensation. In particular, we show that this phase
transition can be associated with an effective chemical potential, at which the
system's bosonic distribution function diverges, and an effective temperature.
As in equilibrium systems, the transition is achieved by tuning this effective
chemical potential to the energy of the lowest normal mode.  Since the
occupations of the modes are available, we determine experimentally observable
properties, such as the luminescence and absorption spectra.
\end{abstract}

\maketitle

\section{Introduction}

Phase transitions in driven-dissipative systems of light strongly coupled to
matter have been the subject of much recent study and progress.
\cite{RevModPhys.85.299} Their intrinsic non-equilibrium nature, caused by the
decay of photons, results in phase transitions that may differ from their
equilibrium analogues.\cite{RevModPhys.85.299, keeling} In many cases, a high
degree of control of the system is possible and questions relating to
non-equilibrium statistical mechanics and the interplay between equilibrium and
non-equilibrium behaviours can be addressed. \cite{PhysRevLett.96.230602,
PhysRevB.75.195331, NanoSciTech.146.Keldysh, FINESSbook,
PhysRevLett.110.195301,
PhysRevA.87.063622} Polaritons confined to two dimensions in semiconductor
microcavities are a well studied example, showing quantum condensation,
\cite{Nature443, Snoke3162007} superfluidity,\cite{Nature457,
amo2009superfluidity, sanvitto2010persistent, PhysRevB.92.035307} and rich
hydrodynamics.\cite{Nature457, nardin2011hydrodynamic, wertz2010spontaneous,
sanvitto2011all, grosso2011soliton, Amo3322011,EPL.110.57006}

Several studies of polaritons have considered the case of incoherent pumping,
where carriers are injected at high energies or momenta and undergo a complex
process of exciton formation and subsequent scattering to relax to the ground
state at the bottom of the lower polariton dispersion.\cite{RevModPhys.85.299}
This scheme was particularly appealing in the search for spontaneous
equilibrium Bose-Einstein Condensation (BEC) of polaritons.\cite{Nature443,
Snoke3162007} However, it leads to a large dynamical exciton reservoir which
affects polariton condensation\cite{PhysRevB.92.035311} and superfluidity. The
complicated pumping and relaxation processes are not well understood, and
models based on phenomenological descriptions suffer from inconsistencies and
divergences.\cite{EPL.102.67007}

Polaritons can also be excited directly by coherent pumping with a
monochromatic laser tuned to the lower polariton dispersion.
\cite{PhysRevLett.84.1547} Under certain pumping conditions, a coherently
pumped polariton system undergoes a phase transition to an optical parametric
oscillator (OPO) regime in which pairs of pump polaritons parametrically
scatter into new signal and idler quasi-condensate states.
\cite{PhysRevLett.85.3680, PhysRevB.62.16247, PhysRevB.65.081308} Without the
exciton reservoir, the theoretical treatment is simplified and an (almost) ab
initio description is possible.\cite{PhysRevB.63.041303, PhysRevB.63.193305,
SemicondSciTech.18.279, PhysRevLett.93.166401, PhysRevB.71.115301,
PSSB:PSSB200560961, PhysRevB.75.075332, PhysRevA.76.043807} In the coherently
pumped configuration we therefore distinguish two distinct regimes: (i) the
regime where only the polariton mode close in energy and momentum to the
external pump is largely occupied, which is the ``coherently pumped'' regime;
(ii) the OPO or ``parametrically pumped'' regime where three modes (the signal,
pump and idler), distinct in energy and momentum, are largely occupied. These
definitions are used throughout this article.

Coherently pumped polariton systems have been used to explore polariton
superfluidity\cite{amo2009superfluidity} since flows with any wave-vector can
be initiated easily. However, under coherent pumping, the phase of the
polariton field is locked to the applied pump, which explicitly breaks the
$U(1)$ symmetry and results in a gapped excitation spectrum.
\cite{PSSB:PSSB200560961} Since this is fundamentally different to what is
expected for a superfluid,\cite{BECPitaevskiiStringari} the interpretation of
the reduced drag when passing an obstacle at small velocity as strong evidence
for a superfluid becomes questionable.

Meanwhile, in the OPO regime, the relative phase of the signal and idler states
is free and the $U(1)$ symmetry is {\it{spontaneously}} broken in each
realisation. The excitation spectrum is now dominated by the gapless Goldstone
mode\cite{PhysRevA.76.043807} exactly as in the incoherently pumped case. Such
a system is expected to be a superfluid according to the traditional
definitions. A recent theoretical and experimental study of flows past an
obstacle in polariton OPO has shown that the coupling between the three OPO
modes leads to rich non-linear behaviour.\cite{PhysRevB.92.035307}

To date, the parametrically pumped polariton system has been described in terms
of dynamical equations, \cite{PhysRevB.71.115301, PSSB:PSSB200560961,
PhysRevLett.93.166401, PhysRevB.75.075332, PhysRevA.76.043807,
SemicondSciTech.18.279, PhysRevB.63.041303, PhysRevB.63.193305} which reflects
that polariton OPO is a purely non-equilibrium phenomenon; in the absence of
drive and decay, no phase transition would exist. The mean-field steady-states
and the excitation spectra both above and below the OPO threshold have been
studied,\cite{PhysRevB.71.115301, PSSB:PSSB200560961, PhysRevB.75.075332,
PhysRevA.76.043807} and quantum Langevin equations have been used to calculate
the photoluminescence below threshold.\cite{SemicondSciTech.18.279,
PhysRevB.63.041303} Recently, an analysis using truncated Wigner methods has
shown that the OPO transition is of the Berezinskii-Kosterlitz-Thouless (BKT)
type and is associated with the binding and proliferation of vortex-
anti-vortex pairs.\cite{PhysRevX.5.041028} The truncated Wigner methods,
however, do not give easy access to correlations at different times and so
energy resolved properties, such as the luminescence and absorption spectra,
are difficult to obtain.

In this article, a quantum field theory for coherently pumped polaritons across
the OPO threshold using Keldysh Green's function techniques, which allows the
calculation of all two time correlation functions including the occupations of
all modes, is developed.\cite{Kamenev, AltlandSimons, KamenevBook} The Keldysh
functional integral approach has been used to study the phase transitions of
several driven-dissipative systems including the superradiant and glassy phase
transitions of the Dicke model,\cite{PhysRevA.87.023831, PhysRevA.87.063622}
BEC of photons in dye-filled cavities,\cite{PhysRevA.88.033829} atoms in
multimode cavities,\cite{PhysRevA.82.043612} and exciton-polaritons in
semiconductor microcavities under incoherent excitation.
\cite{PhysRevLett.96.230602, PhysRevB.75.195331, NanoSciTech.146.Keldysh,
FINESSbook, PhysRevLett.111.026404, NewJPhys.14.065001} It also allows direct
comparison with equilibrium phase transitions.\cite{AltlandSimons}

We find that, despite its highly non-equilibrium nature, the parametrically
pumped polariton system effectively thermalises at low energies and momenta
below the OPO phase transition, and that the approach to the transition is
similar to that in an equilibrium system. In particular, we show that the
system's distribution function diverges at a specific energy, leading to an
effective chemical potential.\cite{NanoSciTech.146.Keldysh, PhysRevB.75.195331}
Moreover, the divergence has the form 1/energy, which leads to the definition
of an effective temperature.\cite{PhysRevA.87.023831} As in equilibrium, the
phase transition is achieved by tuning the effective chemical potential to the
energy of the lowest normal mode.\cite{NanoSciTech.146.Keldysh}

This article is organised as follows: the polariton system and model are
introduced in section \ref{OPOintrod}, and in section \ref{KeldyshBackground}
the Keldysh formalism for the model is set up without making any approximations
about the number of modes. In section \ref{PumptoOPO}, the coherently pumped
system in the pump only, or ``normal'', state is considered and the approach to
the transition discussed in detail, including the concepts of an effective
temperature and chemical potential. The OPO regime, with the
``quasi-condensed'' signal and idler modes, is considered in section
\ref{OPOcalculations}, and conclusions are in section \ref{SummaryConc}.


\section{Polariton System and Model}\label{OPOintrod}

In this section we describe the driven-dissipative polariton system and
introduce a general Hamiltonian.

\subsection{Polaritons}

Microcavity polaritons are the quasi-particles resulting from strong coupling
between quantum well excitons (bound electron-hole pairs) and confined photons
in semiconductor microcavities when the exciton-photon interactions are greater
than the losses.\cite{RevModPhys.85.299,Microcavities} Diagonalising the
Hamiltonian describing a lossless exciton-photon system leads to the new
bosonic eigenstates,\cite{PhysRev.112.1555} the upper and lower polaritons with
dispersions\cite{keeling,RevModPhys.85.299, RevModPhys.82.1489}
$\omega_{up,lp}({\mathbf k}) = \frac{1}{2}\left(\omega_x({\mathbf k}) +
\omega_c({\mathbf k}) \pm \sqrt{(\omega_x({\mathbf k})-\omega_c({\mathbf k}))^2
+ \Omega^2_R}\right),$ where $\omega_x({\mathbf k})$ and
$\omega_c({\mathbf k})$ are the exciton and photon dispersions respectively,
and $\Omega_R$ is the Rabi frequency describing the strength of the coupling.
The polariton operators (upper: $\hat{b}_{\mathbf k}$; lower:
$\hat{a}_{\mathbf k}$) can be written in terms of the exciton,
$\hat{x}_{\mathbf k}$, and photon, $\hat{c}_{\mathbf k}$, operators
\beqy
\hat{a}_{\mathbf k} &=& X({\mathbf k})\hat{x}_{\mathbf k}
+ C({\mathbf k})\hat{c}_{\mathbf k} , \nonumber \\
\hat{b}_{\mathbf k} &=& -C({\mathbf k})\hat{x}_{\mathbf k}
+ X({\mathbf k})\hat{c}_{\mathbf k} , \nonumber \eeqy
where $X({\mathbf k})$ and $C({\mathbf k})$ are the excitonic and photonic
Hopfield coefficients:\cite{RevModPhys.82.1489,SemicondSciTech.18.279}
\beqy
X({\mathbf k}) &=& \frac{\omega_{lp}({\mathbf k}) - \omega_c({\mathbf k})}
{\sqrt{(\omega_{lp}({\mathbf k})-\omega_c({\mathbf k}))^2+
\left(\frac{\Omega_R}{2}\right)^2}} , \label{ExcitonHopfield} \\
C({\mathbf k}) &=& \frac{\Omega_R}{2\sqrt{(\omega_{lp}({\mathbf k})-\omega_c
({\mathbf k}))^2 +\left(\frac{\Omega_R}{2}\right)^2}}, \label{PhotonHopfield}
\eeqy
which are normalised to $X^2({\mathbf k})+C^2({\mathbf k}) = 1$.
$X^2({\mathbf k}),\, C^2({\mathbf k})$ give the exciton and photon fractions of
the lower polaritons respectively.

\begin{figure}[h]
\includegraphics[width=\columnwidth]{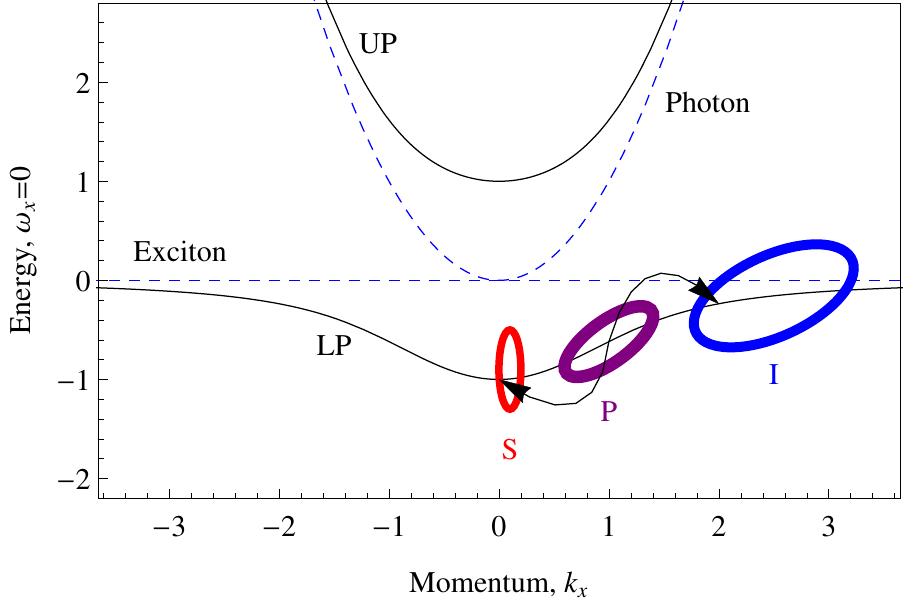}
\caption{(Colour online) Dispersions: UP - upper polariton, LP - lower
polariton, exciton and cavity photon. Pairs of lower polaritons introduced by
an external pump, P (purple), scatter coherently into the signal, S (red), and
idler, I (blue) states while conserving energy and momentum.  OPO with
$k_s \approx 0 $ is observed for a range of pump energies and momenta.
\cite{PhysRevB.68.115325} \label{OPOschematic}}
\end{figure}

The cavity photons are much lighter than the excitons
($m_c \approx 2.3\times10^{-5}m_e, m_x \approx 0.3 m_e$, with $m_e$ the free
electron mass), so the exciton dispersion is usually assumed flat while the
cavity photon dispersion is parabolic $\omega_c({\mathbf k}) = \omega_c(0) +
|{\mathbf k}|^2/2m_c$. The minimum separation of the two polariton branches is
$2\Omega_R$, which occurs at ${\mathbf k}=0$ for $\omega_c(0) = \omega_x$, as
shown in Fig. \ref{OPOschematic}.\cite{keeling, FMMMHSVorticesinOPOChapt,
RevModPhys.82.1489} Different exciton-photon detunings are accessible by choice
of position on the wedge shaped samples used in experiments.
\cite{RevModPhys.82.1489}

As mentioned in the Introduction, we consider creating polaritons by applying a
coherent laser pump near the point of inflection of the lower polariton
dispersion. Above a threshold pump strength, pairs of pump polaritons
spontaneously scatter into new `signal' (${\mathbf k}_s<{\mathbf k}_p,
\omega_s<\omega_p$) and `idler' (${\mathbf k}_i>{\mathbf k}_p,
\omega_i>\omega_p$) states that become extensively occupied. Energy and
momentum are conserved in the scattering process:
$2\omega_p = \omega_s+\omega_i, 2{\mathbf k}_p = {\mathbf k}_s+{\mathbf k}_i$.
This is only possible due to the excitonic non-linearity of the lower
polaritons,\cite{SemicondSciTech.18.279} and is called the optical parametric
oscillator (OPO) regime in analogy to non-linear optics.
\cite{PhysRevB.62.16247} The OPO scattering is shown schematically in
Fig. \ref{OPOschematic}.

For a given set of pump parameters, the observed signal and idler modes are at
well defined energies and momenta,\cite{PhysRevB.68.115325} and have a large
degree of coherence.\cite{PhysRevB.62.16247}

\subsection{Hamiltonian}

We start with a lower polariton Hamiltonian
\beq
\hat{H}_{lp}=\hat{H}_0+\hat{H}_{int}+\hat{H}_{pump}+\hat{H}_{decay},
\label{Hamscheme}
\eeq
describing interacting lower polaritons ($\hat{a}_{\mathbf k},
\hat{a}^\dagger_{\mathbf k}$)
\beqy
\hat{H}_0 &=& \sum_{\mathbf k}\omega_{lp}({\mathbf k})
\hat{a}_{\mathbf k}^\dagger\hat{a}^{}_{\mathbf k}, \nonumber \\
\hat{H}_{int} &=& \sum_{{\mathbf k},{\mathbf k}^\prime,{\mathbf q}}
\frac{V_{{\mathbf k},{\mathbf k}^\prime, {\mathbf q}}}{2}
\hat{a}_{{\mathbf k}}^\dagger\hat{a}_{{\mathbf k}^\prime}^\dagger
\hat{a}^{}_{{\mathbf k}-{\mathbf q}}\hat{a}^{}_{{\mathbf k}^\prime+{\mathbf
q}},
\nonumber
\eeqy
driven by a coherent pump ($F_p$) of the form
\beq
F_p = f e^{-i\omega_p t}e^{i{\mathbf k}_p\cdot {\bf x}}, \label{coherentpump}
\eeq
which introduces polaritons at a single energy, $\omega_p$, and momentum,
${\mathbf k}_p$,
\beqy
\hat{H}_{pump} &=& \hat{a}_{{\mathbf k}_p}^\dagger F_p + F_p^\star
\hat{a}_{{\mathbf k}_p},\label{pumpHam}
\eeqy
and coupled with strength $\Gamma^{\mathbf k}_{\mathbf p}$ to a bosonic decay
bath, $\hat{A}_{\mathbf p}$, with dispersion $\omega^\Gamma_{\mathbf p} =
\omega^\Gamma({\mathbf p})$
\beqy
\hat{H}_{decay} &=& \sum_{{\mathbf k},{\mathbf p}}\Gamma^{\mathbf k}_{\mathbf
p}(\hat{a}_{\mathbf k}^\dagger\hat{A}_{\mathbf p} + \hat{A}^\dagger_{\mathbf
p}\hat{a}_{\mathbf k}) +\sum_{\mathbf p}\omega^\Gamma_{\mathbf
p}\hat{A}_{\mathbf p}^\dagger\hat{A}_{\mathbf p}. \nonumber
\eeqy
Each polariton mode with 2D momentum ${\mathbf k}$ couples to
an independent set of decay bath modes with momenta ${\mathbf p}$,
conserving in plane momentum.\cite{PhysRevB.75.195331} The restriction to
lower polaritons is valid so long as only moderate values of pumping are
used and the pump is always applied close to the lower polariton curve
($\omega_p \approx \omega_{lp}({\mathbf k}_p)$),\cite{PhysRevA.76.043807}
ensuring that non-linear terms involving both polariton branches are
negligible.\cite{PhysRevB.71.115301}

The polariton-polariton interaction strength is given by:
\beqy
V_{{\mathbf k},{\mathbf k}^\prime, {\mathbf q}}& =& g_{\mathrm x}
X({{\mathbf k}})X({\mathbf{k}^\prime})X({{\mathbf k}-{\mathbf q}})X({{\mathbf
k}^\prime+{\mathbf q}}), \label{intstrength}
\eeqy
where the $X({\mathbf k})$'s are defined in Eq. \eqref{ExcitonHopfield} and
reflect that the polariton interactions are due to the excitonic component of
the polaritons. While the exciton-exciton interaction strength $g_{\mathrm x}$
can be assumed constant, the polariton-polariton interaction is in general
momentum dependent. For $\omega_c(0) = \omega_x$, $1/2\leq X^2({\mathbf
k})\leq 1$.\cite{RevModPhys.82.1489}

To proceed further it is helpful to first perform a gauge transformation,
which effectively moves us to the reference frame of the pump mode and makes
the Hamiltonian time-independent. We define the new operator $\tilde{a} =
\hat{a}e^{i\omega_pt}e^{-i{\mathbf k}_p \cdot {\mathbf x}}$ and similarly
$\tilde{a}^\dagger$. Without any loss of generality, we can choose $f\in \Re$
in Eq. \eqref{coherentpump} so the pump term becomes:
\beq
\hat{H}_{pump} \rightarrow f(\tilde{a}^{ }_{{\mathbf 0}} +
\tilde{a}^\dagger_{{\mathbf 0}}). \label{PumppostGT} \nonumber
\eeq
To write the entire Hamiltonian with the new operators ($\tilde{a}$),
note that $\hat{a} = \tilde{a} e^{-i\omega_pt}e^{i{\mathbf
k}_p\cdot{\mathbf x}}$ and that the exponents cancel in all terms
of the Hamiltonian that have the form $\hat{a}^\dagger \hat{a}$ and
$\hat{a}^\dagger\hat{a}^\dagger\hat{a}\hat{a}$. What is left is the momentum
$\mathbf{k}_p$ shift coming from the kinetic energy term.

In the terms that contain the decay bath we now have:
\beqy
\hat{H}_{decay} & = & \sum_{\bf p} \omega^\Gamma_{\mathbf p}\hat{A}_{\mathbf
p}^\dagger \hat{A}_{\mathbf p} + \sum_{{\mathbf k}, {\mathbf p}} \Gamma
(\tilde{a}_{\mathbf k}^\dagger e^{i\omega_pt}e^{-i{\mathbf k}_p \cdot {\mathbf
x}}\hat{A}_{\mathbf p} \nonumber \\
& & \qquad + \hat{A}^\dagger_{\mathbf p}\tilde{a}_{\mathbf k}
e^{-i\omega_pt}e^{i{\mathbf k}_p\cdot{\mathbf x}}). \nonumber
\eeqy
By defining $\hat{A} = \tilde{A}e^{-i\omega_pt}e^{i{\mathbf k}_p \cdot
{\mathbf x}}$ (exactly as the polariton operator), the entire system is
written relative to the pump energy, $\omega_p$, and momentum, ${\mathbf
k}_p$, and the explicit time dependence of Eq. \eqref{pumpHam} is removed.


\section{Keldysh Formalism}\label{KeldyshBackground}

The Keldysh functional integral approach is now applied to both coherently
and parametrically pumped polaritons. In this section, a general complex
Gross-Pitaevskii equation (cGPE) describing the mean field for any ansatz is
obtained without restricting the form of the solution. The process of including
fluctuations is summarised and all quantities of interest are defined.

\subsection{The functional integral representation}

The basis of the functional integral approach is the partition function
of the system that can be written as a coherent state path integral over
bosonic fields $\bar{\psi}$, $\psi$:\cite{Kamenev, KamenevBook, AltlandSimons}
\beq
Z=N\int D(\bar{\psi},\psi)e^{iS}, \nonumber
\eeq
where $N$ provides the correct normalisation and $S$ is the action:
$S=S[\bar{\psi}, \psi]$. In the Keldysh formalism, the system is considered to
evolve from the distant past ($t=-\infty$) to the distant future ($t=+\infty$)
on the forwards branch of a closed time contour and then return along the
backwards branch. The time evolution on the two branches is written in terms
of the separate fields on each branch, $\psi_{f,b}$, and then rotated into
a quantum-classical basis:\cite{Kamenev, KamenevBook, AltlandSimons}
\beq
\psi_{f,b} = \frac{1}{\sqrt{2}}(\psi_{cl}\pm\psi_q).  \label{Keldyshfields}
\eeq
The action can be written as a sum over elements corresponding to the parts
of the Hamiltonian, Eq. \eqref{Hamscheme}:
\beq
iS = i\int d t (S_0 + S_{int} + S_{pump}+S_{decay}). \label{PartsofAction}
\eeq
The components are:
\beqy
&& S_{0} = \sum_{\mathbf k}\Psi_{\mathbf
k}^\dagger(t)(i\partial_t-\omega_{lp}({\mathbf
k+\mathbf{k}_p})+\omega_p)\hat{\sigma}_1^K\Psi_{\mathbf k}(t)
\label{polaritonsaction}, \\
&& S_{int} = -\sum_{{\mathbf k},{\mathbf k}^\prime,{\mathbf
q}}\frac{V_{{\mathbf k},{\mathbf k}^\prime, {\mathbf q}}}{4}
\big(\Psi^\dagger_{\mathbf k}(t)\hat{\sigma}^K_1\Psi_{{\mathbf k}-{\mathbf
q}}(t)\Psi^\dagger_{{\mathbf k}^\prime}(t)\Psi_{{\mathbf k}^\prime +{\mathbf
q}}(t) \nonumber \\
&& \hspace{1.2cm} + \Psi^\dagger_{{\mathbf k}}(t)\Psi_{{\mathbf k}-{\mathbf
q}}(t)\Psi^\dagger_{{\mathbf k}^\prime}(t)\hat{\sigma}^K_1\Psi_{{\mathbf
k}^\prime + {\mathbf q}}(t)\big) \label{interactionaction}, \\
&& S_{pump} = -\sqrt{2}f(\bar{\psi}_{{\mathbf 0},q}(t)+\psi_{{\mathbf 0},q}(t))
\label{pumpaction},\\
&& S_{decay} = \sum_{\mathbf p}\chi^\dagger_{\mathbf p}(t)(i\partial_t -
\omega^\Gamma_{{\mathbf{p +k}_p}}+\omega_p)\hat{\sigma}_1^K\chi_{\mathbf p}(t)
\label{originaldecayaction} \\
&& \hspace{1.2cm}- \sum_{{\mathbf k},{\mathbf p}}\Gamma^{\mathbf k}_{\mathbf
p}(\chi^\dagger_{\mathbf p}(t)\hat{\sigma}_1^K\Psi_{\mathbf k}(t) +
\Psi^\dagger_{\mathbf k}(t)\hat{\sigma}_1^K\chi_{\mathbf p}(t)), \nonumber
\eeqy
where $\hat{\sigma}_1^K$ is the Pauli matrix\cite{LandauV3} in the Keldysh
quantum-classical space and the bath, $\chi$, and polariton, $\Psi$, fields
are written as vectors of the quantum and classical fields:
\beqs
\hat{\sigma}^K_1 = \bpm 0&1\\1&0 \epm, \Psi_{\mathbf k}(t) = \bpm
\psi_{{\mathbf k},cl}(t) \\ \psi_{{\mathbf k},q}(t) \epm, \chi_{\mathbf p}(t)
= \bpm \chi_{{\mathbf p},cl}(t) \\ \chi_{{\mathbf p},q}(t) \epm.
\eeqs
Since the pump is classical, i.e. the same on the forward and backward branches
of the Keldysh contour, after the Keldysh rotation it only couples to quantum
fields in Eq. \eqref{pumpaction}. Moreover, the pump injects polaritons with
momentum $\mathbf{k}_p$ which, after the gauge transformation to the pump
frame, corresponds to zero momentum in Eq. \eqref{pumpaction}. The momentum
${\mathbf{k}}_p$ and energy $\omega_p$ shifts in Eqs. \eqref{polaritonsaction}
and \eqref{originaldecayaction} come from the spatial (Fourier transform of
the polariton and the bath's dispersion operators) and time derivatives acting
on $\hat{a}_{\mathbf{k}}=\tilde{a}_{\mathbf{k}}e^{-i\omega_pt}e^{i{\mathbf
k}_p\cdot{\mathbf x}}$ and $\hat{A} = \tilde{A}e^{-i\omega_pt}e^{i{\mathbf
k}_p \cdot {\mathbf x}}$ respectively.

\subsection{Integrating out the decay bath} \label{decaybath}

The bath fields are present in the action at quadratic level so the functional
integral over them can be performed analytically.\cite{PhysRevB.75.195331,
PhysRevA.82.043612} The procedure is the same as for the photon decay bath in
earlier studies of driven-dissipative polaritons.\cite{PhysRevB.75.195331}
After performing the Gaussian integration over the decay bath's fields, we
Fourier transform into the energy representation and make a series of standard
assumptions about the properties of the decay bath.\cite{PhysRevB.75.195331}

In particular, it is assumed that the bath couples equally to all polariton
modes $\hat{a}_{\mathbf k}$, $\Gamma^{\mathbf k}_{\mathbf p} \rightarrow
\Gamma_{\mathbf p}$, making the polariton decay momentum independent. To
include full momentum dependence in the polariton decay, one would either
have to explicitly include momentum dependent coupling here, presupposing
knowledge of the correct Hopfield coefficient weightings, or consider two
independent baths of excitons and photons which would be weighted properly
as a result of the rotation from the exciton-photon basis.

Further, the decay bath is assumed to be large and unaffected by the behaviour
of the system, that the coupling $\Gamma_{\mathbf p}$ is a smooth function of
${\mathbf p}$, and that the bath has a dense energy spectrum, so the summation
over ${\mathbf p}$ can be replaced by $\int d\omega^\Gamma$. Although any form
of the decay bath's density of states and coupling to the polariton system
could be chosen, it is reasonable to assume that they are constant. This leads
to self-energy contributions from the decay bath:\cite{PhysRevB.75.195331}
\beqy
d^{R,A}(\omega)&=&\mp i\kappa_{lp},\nonumber\\
d^K(\omega)&=&-2 i\kappa_{lp} F_{\chi}(\omega+\omega_p)\nonumber,
\eeqy
where $\kappa_{lp}$ is the constant polariton decay rate and $F_{\chi}(\omega)$
is the bath's distribution function. The presence of $\omega_p$ is due to
gauge transformation to the pump frame. Note that since there is still an
explicit $\omega$ dependence in the bath distribution present in the Keldysh
part of the self-energy, the influence of the bath is non-Markovian at this
stage. However, as discussed in detail in Ref. \onlinecite{PhysRevA.87.023831},
since the drive frequency $\omega_p$ is much larger then the energy associated
with the room temperature thermal photons outside the cavity, decay bath modes
with energies in the range of interest $|\omega| < 2\Omega_R$ are effectively
not occupied, and $F_{\chi}(\omega+\omega_p)$ can be set to 1. However, if
needed, inclusion of a frequency dependent decay bath in both the retarded
(deterministic in the case of Wigner approach \cite{PhysRevB.72.125335}) and
Keldysh (stochastic) components is straightforward in the Keldysh formalism.

The decay term in the Keldysh action now contains only the polariton fields,
and after inverting the Fourier transform, Eq. \eqref{originaldecayaction}
is replaced by:
\beqy
 S_{decay} &=& \kappa_{lp}\sum_{\mathbf k}\int d t \Psi^\dagger_{\mathbf
 k}(t)\hat{\sigma}^K_2\Psi_{\mathbf k} (t) \label{finaldecayaction} \\
 && + 2i\kappa_{lp}\sum_{\mathbf k}\int dt dt^\prime \bar{\psi}_{{\mathbf
 k},q}(t)F_\chi(t-t^\prime)\psi_{{\mathbf k},q}(t^\prime), \nonumber
 \eeqy
where $\hat{\sigma}^K_2$ is the Pauli matrix
\beq
\hat{\sigma}^K_2 = \bpm 0 & -i \\ i & 0 \epm. \nonumber
\eeq

\subsection{Mean field and saddle points}\label{MFSPsbackground}

The mean field equation is calculated from the saddle points of the
action\cite{Kamenev, AltlandSimons, PhysRevB.75.195331} taken relative to
both the classical and quantum fields. This involves finding the solutions to
\beqs
\frac{\partial S}{\partial\bar{\psi}_{{\mathbf k},cl}(t)}=0 \qquad \mathrm{and}
\qquad \frac{\partial S}{\partial \bar{\psi}_{{\mathbf k}, q}(t)}=0,
\eeqs
which leads to
\beqy
\frac{\partial S}{\partial \bar{\psi}_{{\mathbf k},
cl}(t)} &=& (i\partial_t-\omega_{lp} ({\mathbf k}+{\mathbf
k}_p)+\omega_p-i\kappa_{lp})\psi_{{\mathbf k},q}(t) \nonumber \\
&& -\sum_{{\mathbf k}^\prime,{\mathbf q}}\frac{V_{{\mathbf k},{\mathbf
k}^\prime,{\mathbf q}}}{2} \Big( \bar{\psi}_{{\mathbf k}^\prime,
q}(t)\Psi^T_{{\mathbf k}-{\bf q}}(t)\Psi_{{\mathbf k}^\prime+{\mathbf q}}(t)
\nonumber \\
&& \hspace{1.5cm} + \bar{\psi}_{{\mathbf k}^\prime,cl}(t)\Psi^T_{{\mathbf
k}-{\mathbf q}}(t)\hat{\sigma}^K_1\Psi_{{\mathbf k}^\prime+{\mathbf
q}}(t)\Big)\nonumber
\eeqy
and
\beqy
\frac{\partial S}{\partial \bar{\psi}_{{\mathbf k},q}(t)} &=&
(i\partial_t-\omega_{lp} ({\mathbf k} + \mathbf{k}_p)+\omega_p +
i\kappa_{lp})\psi_{{\mathbf k}, cl}(t) \nonumber \\
&& -\sum_{{\mathbf k}^\prime,{\mathbf q}}\frac{V_{{\mathbf k},{\mathbf
k}^\prime, {\mathbf q}}}{2} \Big( \bar{\psi}_{{\mathbf k}^\prime,
q}(t)\Psi^T_{{\mathbf k}-{\mathbf q}}(t)\hat{\sigma}^K_1\Psi_{{\mathbf
k}^\prime+{\mathbf q}}(t)\nonumber \\
&& \hspace{1.9cm} + \bar{\psi}_{{\mathbf k}^\prime,cl}(t)\Psi^T_{{\mathbf
k}-{\mathbf q}}(t)\Psi_{{\mathbf k}^\prime + {\mathbf q}}(t)\Big) \nonumber \\
&& -\sqrt{2}f\delta_{{\mathbf k},0} -2i\kappa_{lp} \int dt^\prime
F_\chi(t-t^\prime)\psi_{{\mathbf k},q}(t^\prime). \nonumber
\eeqy

There always exists a solution to the saddle point equations where the
quantum part is zero, which corresponds to the purely classical solution
and leads to the complex Gross-Pitaevskii equation (cGPE) of the mean-field
analysis.\cite{PhysRevB.75.195331, Kamenev,AltlandSimons} With $\psi_{\mathbf
k,q}(t) = 0$, $\partial S/\partial \bar{\psi}_{\mathbf k, cl}(t)=0$ is
automatically satisfied, and the second equation becomes
\beqy
&& \hspace{-0.4cm} \frac{\partial S}{\partial \bar{\psi}_{{\mathbf
k},q}(t)}= (i\partial_t -\omega_{lp} ({\mathbf
k}+\mathbf{k}_p)+\omega_p+i\kappa_{lp})\psi_{{\mathbf k},cl}(t) \nonumber \\
&& \hspace{0.2cm} - \sqrt{2}f\delta_{{\mathbf k},0} -\sum_{{\mathbf
k}^\prime,{\mathbf q}}\frac{V_{{\mathbf k},{\mathbf k}^\prime, {\mathbf
q}}}{2} \bar{\psi}_{{\mathbf k}^\prime,cl}(t)\psi_{{\mathbf k}-{\mathbf
q},cl}(t)\psi_{{\mathbf k}^\prime+ {\mathbf q},cl}(t). \nonumber
\eeqy
$\psi_{cl}$ at the saddle point, $\psi^{sp}_{cl}$, is $\sqrt{2}$
times the mean field value $\psi^{\mathrm{mf}}$ as defined in other
approximations\cite{PhysRevB.71.115301, PhysRevB.75.075332}. The
mean-field polariton field is equal to the $\psi_f$, as defined
by Eq. \eqref{Keldyshfields}, at the saddle point.\cite{Kamenev,
PhysRevB.75.195331} We therefore have:
\beqy
0&=&(i\partial_t -\omega_{lp}({\mathbf k}+\mathbf{k}_p) +\omega_p
+i\kappa_{lp})\psi_{\mathbf k}^{\mathrm{mf}} - f\delta_{{\mathbf k},0}
\nonumber \\
&& -\sum_{{\mathbf k}^\prime,{\mathbf q}}V_{{\mathbf k},{\mathbf k}^\prime,
{\mathbf q}}{\bar{\psi}_{{\mathbf k}^\prime}^{\mathrm{mf}}}\psi_{{\mathbf
k}-{\mathbf q}}^{\mathrm{mf}}\psi_{{\mathbf k}^\prime+{\mathbf
q}}^{\mathrm{mf}} \label{generalcGPE}.
\eeqy
As will be seen, by restricting to the appropriate modes, the usual cGPEs
describing the polariton OPO \cite{PhysRevB.71.115301, PhysRevB.75.075332}
are reproduced. In this article we consider steady-state solutions to
Eq. \eqref{generalcGPE} in the coherently pumped and OPO regimes.

\subsection{Fluctuations about the mean field}\label{generalfluctuations}

Having found the mean-field cGPEs and their steady-state solutions,
the next task is to consider whether the solutions of these equations
are stable to small fluctuations close in energy and momentum to the
mean-field state. \cite{PhysRevA.76.043807,PhysRevB.71.115301} A physical
solution requires stability of the saddle point equations so only if
the state is stable can other quantities such as the luminescence be
calculated.\cite{PhysRevB.75.195331,PhysRevB.76.115326} We therefore add
small fluctuations $\delta \psi_{cl}$ and $\delta\psi_q$ to the mean field
and construct the inverse Green's functions by substituting
\beq
\psi_{cl} = \psi^{sp}_{cl} + \delta \psi_{cl}
=\sqrt{2}\psi^{\mathrm{mf}}+\delta\psi_{cl} \label{clfluctuations}
\eeq
and
\beq
\psi_{q} = \delta\psi_q \label{qfluctuations}
\eeq
into
Eqs. \eqref{polaritonsaction}-\eqref{pumpaction} and \eqref{finaldecayaction},
and restricting to terms that are second order in the fluctuations. Since
all quantities of interest will be in energy-momentum space, the Fourier
transform is performed at the level of the action of fluctuations. The
remaining part of the action then has the form:
\beq
S[\Delta\Psi] = \int d\omega \sum_{\mathbf{k}}
\Delta\Psi_\mathbf{k}^\dagger(\omega) \bpm 0 & [D^{-1}]^A \\ [D^{-1}]^R &
[D^{-1}]^K \epm \Delta\Psi_{\mathbf{k}}(\omega) \label{fluctuationActionform}
\eeq
where $[D^{-1}]^{\{R/A/K\}}=[D^{-1}]^{\{R/A/K\}}(\omega, \mathbf{k})$ and $R,
A, K$ indicate the retarded, advanced and Keldysh components of the inverse
Green's function respectively. The fluctuations are written using a Nambu
vector form: \cite{PhysRevB.75.195331}
\beq
\Delta\Psi_{\mathbf{k}}(\omega) = \bpm \delta\psi_{\mathbf{k}, cl}(\omega)
\\ \delta\bar{\psi}_{-\mathbf{k}, cl}(-\omega) \\ \delta\psi_{\mathbf{k},
q}(\omega) \\ \delta\bar{\psi}_{-\mathbf{k}, q}(-\omega)\epm, \nonumber
\eeq
where the fluctuations are relative to the pump mode. If there
is more than one mode, each $\delta\psi$ has the structure
$(\delta\psi_{m1}, \delta\psi_{m2} \ldots)^T $ and the fluctuations
are relative to each mode ($_{\pm\mathbf{k}}(\pm\omega) \rightarrow
_{\mathbf{k}_m\pm\mathbf{k}}(\omega_m\pm\omega)$).  Further, the following
relations are used:\cite{Kamenev, KamenevBook}
\beqy
[D^{-1}]^R &=& {[D^R]}^{-1} \label{DRinversion},\\
D^A &=& {\left(D^R\right)}^\dagger , \label{DArelDR} \\
{D^K = -D^R [D^{-1}]^K D^A} &=& D^RF_s - F_s D^A\label{DKinversion},
\eeqy
where $F_s$, Fourier transformed into energy, is the distribution matrix
of the system.\cite{Kamenev} The poles of the retarded Green's function,
$\omega^{\pm}$, give the spectrum of excitations, while the signs of their
imaginary parts determine whether the proposed mean-field steady state is
stable (a positive imaginary part, $\Im(\omega^{\pm})>0$, implies that the
proposed steady state is unstable). Solving $\det([D^{-1}]^R) =0$ for complex
$\omega$ is equivalent to calculating the eigenvalues in linear response
analysis.\cite{PhysRevA.76.043807, PSSB:PSSB200560961}

To calculate physical observables such as the luminescence and absorption
spectra of the polariton system, the Keldysh rotation is inverted to find
the forwards $(<)$ and backwards $(>)$ Green's functions:\cite{Kamenev,
KamenevBook, PhysRevB.75.195331}
\beqy
D^{<,>}&=&\frac{1}{2}(D^K\mp(D^R-D^A)); \label{dlessgreat}\\
D^< &=& -i\langle \psi_f \psi_b^\dagger \rangle. \nonumber
\eeqy
In energy ($\omega$) - momentum (${\mathbf k}$) space, these give the
incoherent luminescence
\beq
L(\omega, {\mathbf k})= \frac{i}{2\pi}D^{<}(\omega, {\mathbf k}),
\label{lumdef}
\eeq
and absorption
\beq
A(\omega, {\mathbf k})=\frac{i}{2\pi}D^{>}(\omega, {\mathbf k}) \label{absdef}
\eeq
spectra in the steady-state.\cite{PhysRevB.75.195331} Note that since
$D^{R, A, K}$ come from inverting Eq. \eqref{fluctuationActionform}
using Eqs. \eqref{DRinversion}-\eqref{DKinversion}, they all contain
$\det([D^{-1}]^R)$ in the denominator; if this is zero while the numerators
of $D^{<}$ remain non-zero, then the luminescence diverges, indicating an
instability of the chosen mean-field solution towards a transition to some
different phase.

In experiments polaritons are observed through the photon losses from the
microcavities. The photon luminescence is obtained by extracting the photon
part through multiplication by the (momentum dependent) photon fraction
$C^2({\mathbf k})$, defined in Eq. \eqref{PhotonHopfield}:
\beq
L_{phot} = C^2({\mathbf k})L_{lp}. \label{PhotRescale}
\eeq
The spectral weight is defined as the difference between the absorption and
luminescence. In terms of the Green's functions\cite{PhysRevB.76.115326},
\beq
SW(\omega, {\mathbf k}) = A(\omega, {\mathbf k}) -L(\omega, {\mathbf k}) =
\frac{i}{2\pi}(D^R-D^A), \label{swdef}
\eeq
or $1/(2\pi)$ times the spectral response of
Ref. \onlinecite{PhysRevA.87.023831}.

It should further be noted that recent work on the incoherently pumped
polariton system, that considers the phase fluctuations to all orders
with Renormalisation Group analysis, has revealed that the long-wavelength,
long-time behaviour depends strongly on the anisotropy.\cite{PhysRevX.5.011017}
In particular, the isotropic limit of an infinite system has been shown to
require an exact treatment of phase fluctuations, but a certain degree of
spatial anisotropy leads the system to fall into the usual Gaussian fixed
point, such as that considered in this article. Our coherently pumped polariton
system is intrinsically strongly anisotropic with the polariton dispersion
substantially different along the two spatial directions perpendicular and
parallel to the pump wave-vector. This suggests that the Gaussian description
will be sufficient for studying parametrically pumped polaritons in certain
regimes, and is certainly sufficient for the coherently pumped polaritons,
where the Goldstone mode and so large phase fluctuations are absent.

\subsection{Dimensionless units}

Throughout this paper, a flat exciton dispersion $\omega_x(\mathbf{k}) =
\omega_x$ is used. We work in a system of non-dimensional units where all
the energies are rescaled such that $\Omega_R/2 = 1$, and all the momenta or
wave-vectors (with $\hbar=k_B = 1$) are rescaled to ${\bf k} \rightarrow {\bf
k}/\sqrt{\Omega_rm_c}$.  The key control parameters are the pump wave-vector,
which for OPO is typically in the range of\cite{PhysRevB.68.115325,
PhysRevB.70.205301} $1.0 \mu \mathrm{m}^{-1} <|\mathbf{k}_p| < 2.0\mu
\mathrm{m}^{-1}$, and the detuning of the pump energy away from the polariton
dispersion $\omega_p - \omega_{lp}(\mathbf{k}_p)$. In particular for all
results presented in this article, $\mathbf{k}_p = (1.5,0)$ in dimensionless
units along the x-direction is used, which corresponds to $|\mathbf{k}_p|
\approx 1.677\mu \mathrm{m}^{-1}$ for realistic $\Omega_R = 5 \mathrm{meV}$
and $m_c = 0.25\times 10^{-5} m_e$. \cite{PhysRevB.75.075332,keeling}
The pump is chosen to be resonant with the lower polariton dispersion at
the pump momentum, $\omega_p - \omega_{lp}(\mathbf{k}_p) = 0$. This means
that bistability in the pump mode is avoided by working in the so called
`optical limiter' regime. \cite{PhysRevA.69.023809, PhysRevB.75.075332,
PSSB:PSSB200560961} With $\omega_x = 0$, for $\mathbf{k}_p=(1.5,0)$ we get
$\omega_p = -0.380$.  The polariton decay $\kappa_{lp}$ we choose to be $
0.05$ in dimensionless units which corresponds to $\kappa_{lp} = 0.125
\mathrm{meV}$. The exciton-exciton interaction strength $g_{\mathrm x}$
is implicitly rescaled to $1$ which in turn applies a scaling to the fields.


\section{Transition to OPO} \label{PumptoOPO}

As in previous theoretical studies of the OPO
transition \cite{PhysRevB.63.041303, PhysRevB.75.075332,
PSSB:PSSB200560961,PhysRevB.71.115301} we consider a single mode mean-field
solution first i.e assuming that only the mode with frequency and wave
vector equal to that of the external pump is occupied, and then investigate
the appearance of the additional signal ($\mathbf{k}_s<\mathbf{k}_p,
\omega_s<\omega_p$) and idler ($\mathbf{k}_i>\mathbf{k}_p, \omega_i>\omega_p$)
modes with momenta and energies determined by momentum and energy
conservation. The method described in the previous section is applied to this
`normal state' of coherently pumped polaritons.

\subsection{Mean field}

To get the simplest single-mode mean-field solution we start from
Eq. \eqref{generalcGPE}, make a plane wave ansatz for the solution at the
pump frequency and wave vector, restrict the interactions to remain within
the pump mode, and calculate the steady state. The plane wave for the pump
mode mean-field in the rotating frame is just a constant, complex, amplitude:
\beq
\psi_p^{\mathrm{mf}} = P . \nonumber
\eeq
In the steady state, $\partial_t P = 0$, this gives:\cite{PhysRevB.75.075332,
PhysRevB.71.115301}
\beq
0 = (\omega_{lp}({\mathbf k}_p)-\omega_p - i\kappa_{lp} + V_{ppp}|P|^2)P +
f. \label{pumpmeanfield}
\eeq
Note that from Eq. \eqref{pumpmeanfield} onwards, instead of writing the
momenta in the interaction (Eq. \eqref{intstrength}), the mode (here $p$)
index and the fluctuation signature $m^\pm$ are used as appropriate, where
$m$ is any mode.

Taking the modulus square of Eq. \eqref{pumpmeanfield}, the mode occupation
$n_p = |P|^2$ is related to the pump strength $I_p=|f|^2$ as follows:
\beq
I_p = (\omega_{lp}({\mathbf k}_p)-\omega_p +V_{ppp}n_p)^2n_p + \kappa_{lp}^2n_p
.\label{pumppower}
\eeq
Without any loss of generality, the pumping field amplitude can be chosen
to be real ($f = \sqrt{I_p}$). A weak pump corresponds to a low $n_p$ while
a strong pump has a high $n_p$. A typical behaviour of $n_p$ as a function
of pump power $I_p$ is shown in Fig. \ref{OPOmf} (the region where $n_s$
is zero corresponds to the pump only state
discussed in this section).

\subsection{Green's functions and distribution matrix}\label{divergencessect}

The inverse Green's functions are obtained by considering small fluctuations
about the mean field. The pump mode with fluctuations can be written in
the quantum-classical basis in the form of Eqs. \eqref{clfluctuations}
and \eqref{qfluctuations} as follows:
\beq
\psi_{\mathbf{k}, cl}(\omega) =
\sqrt{2}P\delta_{\mathbf{k},\mathbf{0}}\delta_{\omega,0} + \delta
\psi_{\mathbf{k}, cl} (\omega), \label{pumpclfluct}
\eeq
\beq
\psi_{\mathbf{k},q} = \delta \psi_{\mathbf{k},q}(\omega), \label{pumpqfluct}
\eeq
where the $\sqrt{2}$ in the classical part comes from the mean field solution
being on the forwards branch of the closed time contour. As described in
section \ref{generalfluctuations}, only terms second order in the fluctuations
are kept and there is an implicit summation over momenta. In the pump only
state, the inverse retarded Green's functions in \eqref{fluctuationActionform}
are:
\beq
[D^{-1}]^R(\omega,\mathbf{k}) = \frac{1}{2} \bpm \omega-\alpha^+ +i\kappa_{lp}
& -V_{p_{+-}}P^2 \\ -V_{p_{+-}}{P^*}^2 & -\omega-\alpha^--i\kappa_{lp}
\epm \quad \label{PumpDm1R}
\eeq
where $\alpha^\pm = \omega_{lp}({\mathbf k}_p\pm
\mathbf{k})-\omega_p+2V_{p_\pm{}_\pm} n_p$, and
\beq
[D^{-1}]^K(\omega,\mathbf{k}) = i\kappa_{lp}\bpm F_\chi(\omega+\omega_p) &
0 \\0 & F_\chi(-\omega-\omega_p)\epm. \quad \label{PumpDm1K}
\eeq

Although the bath's distribution $F_\chi(\omega)$ can have any form, we
choose $F_\chi (\epsilon) = 2n_B(\epsilon) +1 $, where $n_B(\epsilon)$
is the Bose-Einstein distribution, to represent thermal modes outside of
the cavity. As discussed in section \ref{decaybath} $F_\chi(\pm\omega\pm
\omega_p) \approx 1$. Using Eqs. \eqref{DRinversion}-\eqref{DKinversion}
and \eqref{PumpDm1R}-\eqref{PumpDm1K},
\beq
D^R(\omega,\mathbf{k}) = A\bpm -\omega - \alpha^- -i\kappa_{lp} & V_{p_{+-}}P^2
\\ V_{p_{+-}}{P^*}^2 & \omega - \alpha^+ + i\kappa_{lp} \epm, \nonumber
\eeq
\beq
D^K(\omega,\mathbf{k}) = B \bpm \nu^-& \eta^- P^2 \\ \eta^+{P^*}^2 & \nu^+
\epm, \nonumber
\eeq
where the following shorthand notation has been introduced
\beqy
\nu^\pm &=& (\omega\mp\alpha^{\pm})^2 + \kappa_{lp}^2 + V^2_{p_{+-}}n^2_p,
\nonumber \\
\eta^\pm &=& (-(\alpha^++\alpha^-)\pm2i\kappa_{lp})V_{p_{+-}}, \nonumber
\eeqy
and the pre-factors are:
\beqy
A &=& \frac{1}{2\det([D^{-1}]^R(\omega, \mathbf{k}))}, \nonumber \\
B &=& \frac{-i\kappa_{lp}}{4{|\det([D^{-1}]^R(\omega, \mathbf{k}))|}^2}.
\nonumber
\eeqy
With the simple matrix structure, Eq. \eqref{DKinversion} can be used to
find the distribution matrix:
\beq
F_s(\omega, \mathbf{k}) = \bpm 1 & \frac{-2V_{p_{+-}}P^2}{2\omega -\alpha^+
+\alpha^-}\\ \frac{-2V_{p_{+-}}{P^*}^2}{2\omega -\alpha^+ +\alpha^-} &
-1 \epm. \label{pumpFsmatrix}
\eeq
When $\omega-(\alpha^+-\alpha^-)/2=0$, the distribution matrix
diverges. Although $F_s$ can in general be very different to the equilibrium
Bose-Einstein distribution, by comparing the two, an effective chemical
potential can be introduced as the energy at which the distribution function
diverges:
\beq
\mu_{\mathrm{eff}}(\mathbf{k}) = \frac{\alpha^+(\mathbf{k})
-\alpha^-(\mathbf{k})}{2}.\label{pumpchempot}
\eeq

\subsection{Zeros of $[D^{-1}]^R$ - spectra and chemical potential}

The inverse retarded Green's function $[D^{-1}]^R$ is related to
the Bogoliubov matrix $L$ obtained by considering small fluctuations
about the steady state.\cite{PhysRevB.75.075332, PhysRevB.63.041303,
PSSB:PSSB200560961} In particular, $L(\mathbf{k}) = -2\sigma_z ([D^{-1}]^R
(0,-\mathbf k))$, and solving $\det([D^{-1}]^R(\omega^{\pm},\mathbf{k}))
=0 $ for $\omega^{\pm}(\mathbf{k}) \in \C$ is equivalent to finding the
eigenvalues of $L$:\cite{PhysRevB.75.195331, PhysRevB.75.075332}
\beq
\omega^{\pm}(\mathbf{k}) = \frac{\alpha^+-\alpha^-}{2}
-i\kappa_{lp} \pm \frac{1}{2}\sqrt{(\alpha^++\alpha^-)^2
-4V_{p_{+-}}^2n_p^2)}. \label{pumpeigens}
\eeq
The real parts of $\omega^{\pm}(\mathbf{k})$ ($\Re(\omega^{\pm}(\mathbf{k}))$)
correspond to the spectra of excitations while the signs of the imaginary
parts of $\omega^{\pm}(\mathbf{k})$ determine whether the proposed mean-field
solution is stable (if $\Im(\omega^{\pm})>0$ at any momentum, the proposed
single-mode solution is unstable).

\begin{figure}[h]
\includegraphics[width=\columnwidth]{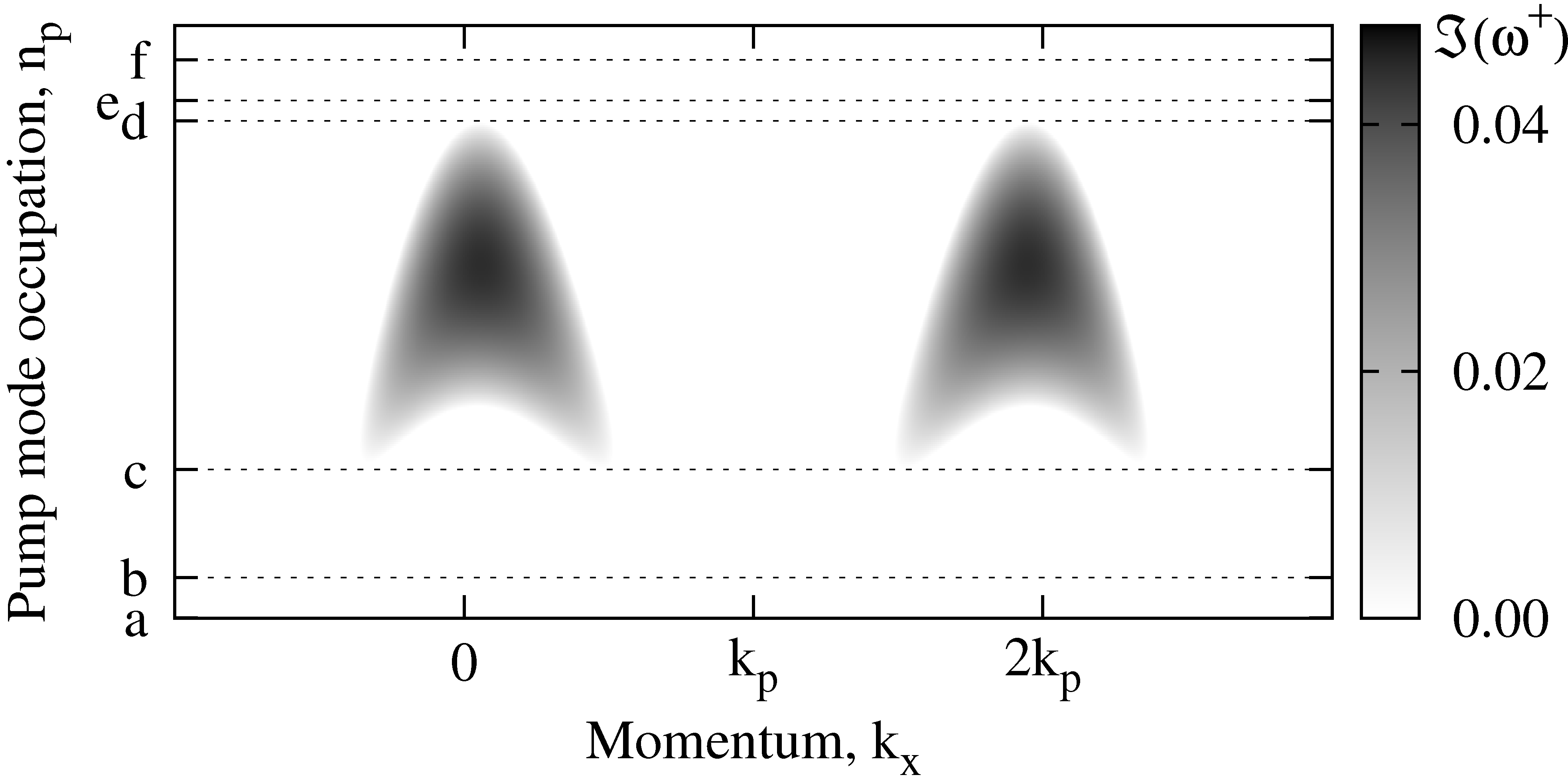}
\caption{Regions of unstable single-mode pump-only state,
$\Im(\omega^{\pm})>0$, symmetric around the pump momentum $\mathbf{k}_p$. The
letters and dotted lines correspond to the pump mode occupations used in
figure \ref{simultaneouszeros} (a is $n_p = 1\times10^{-4}$ and cannot be
resolved from the horizontal axis). \label{PumpUnstable}}
\end{figure}

The unstable region depends on the external drive which is related to the pump
mode occupation by Eq. \eqref{pumppower}. After transforming back to the lab
frame, Fig. \ref{PumpUnstable} shows that there are two regions around $k_x=0$
and $k_x=2|\mathbf{k}_p|$ where the single mode becomes unstable to small
fluctuations as the external drive strength is increased (decreased) above
(below) some lower (upper) threshold value. For the parameters chosen, the
stability diagram presented in Fig. \ref{PumpUnstable} shows the instability
towards an OPO state. It also demonstrates that there is no bistability in
the pump occupation since the instabilities occur away from the pump momentum.

It is also instructive to examine the real and imaginary parts of
$\det([D^{-1}]^R(\omega, \mathbf{k}))$. Considering
$\Re[\det([D^{-1}]^R(\xi^{\pm},\mathbf{k}))] = 0$ gives:
\beq
\xi^{\pm}(\mathbf{k}) = \frac{\alpha^+ -\alpha^-}{2} \pm
\frac{1}{2}\sqrt{(\alpha^++\alpha^-)^2 +4(\kappa_{lp}^2 -V_{p_{+-}}^2n_p^2)}
\label{pumppoles}
\eeq
while the requirement for the imaginary part of the determinant of the retarded
inverse Green's function being zero corresponds to the same condition as
the divergence of the distribution $F_s$ discussed in previous section
i.e. $\Im[\det([D^{-1}]^R(\mu_{\mathrm{eff}},\mathbf{k}))] = 0$.

\begin{center}
\begin{figure*}[h]
\includegraphics[width=\columnwidth]{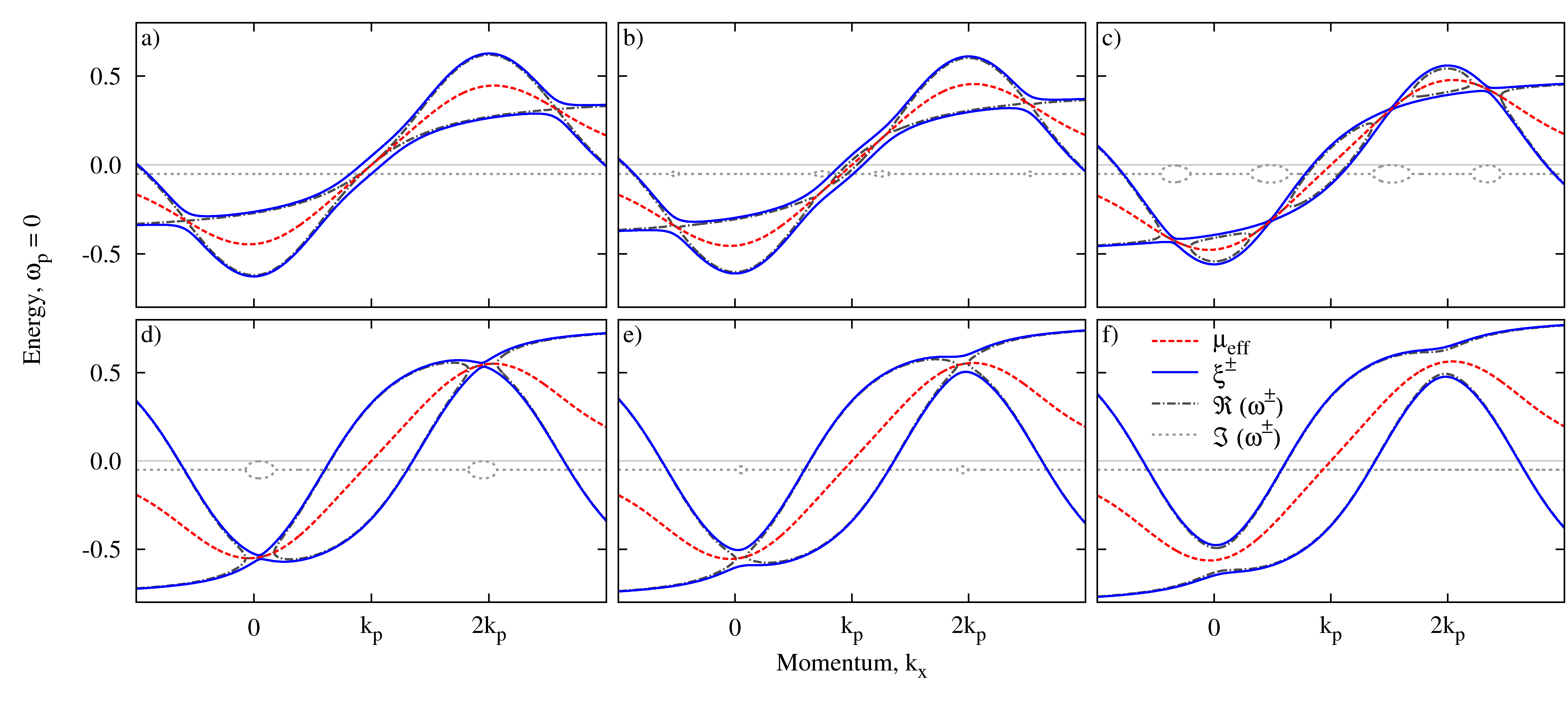}
\caption[]{(Colour online) Solutions to $\det([D^{-1}]^R)=0$ for the
  stable pump mode occupations indicated in
  Fig. \ref{PumpUnstable}. Red dotted: $\mu_{\mathrm{eff}}$ from
  $\Im(\det([D^{-1}]^R(\mu_{\mathrm{eff}},\mathbf{k})))=0$ where
  $\mu_{\mathrm{eff}} \in \R$ ; solid blue: $\xi^\pm$ from
  $\Re(\det([D^{-1}]^R(\xi^\pm,\mathbf{k})))=0$ where $\xi^\pm \in
  \R$; dark grey dashed: $\Re(\omega^\pm)$ and grey dashed:
  $\Im(\omega^\pm)$ from $\det([D^{-1}]^R(\omega^{\pm},\mathbf{k}))=0$
  where $\omega^{\pm} \in \C$. Top row: approaching lower threshold
  from below: a) $n_p = 1\times 10^{-4}$; b) $n_p = 0.02$; c) $n_p =
  0.073$. Bottom row: increasing $n_p$ above `upper threshold': d)
  $n_p=0.245$; e) $n_p = 0.255$; f) $n_p =
  0.275$. \label{simultaneouszeros}}
\end{figure*}
\end{center}

In Fig. \ref{simultaneouszeros}, $\Re(\omega^\pm(\mathbf{k})),
\Im(\omega^\pm(\mathbf{k})), \xi^\pm(\mathbf{k})$ and
$\mu_{\mathrm{eff}}(\mathbf{k})$ are plotted for a range of stable pump
mode occupations. First the case of a weak pump (low $n_p$) is considered
and the instability threshold is approached from below, as shown in the
top row of Fig. \ref{simultaneouszeros}. The imaginary parts of the complex
eigenvalues $\Im(\omega^\pm(\mathbf{k}))$ start to split and the real parts
combine in four distinct regions, leading to the double tails when the pump
state first becomes unstable, as seen in Fig. \ref{PumpUnstable}. When the
four maxima in the imaginary parts of the complex eigenvalues first appear,
two are located near to the pump momentum, one at a much higher and one at
a much lower momentum. As the transition is approached, their values grow
and those that were below the pump momentum move towards each other and the
$k_x=0$ point, while those that were above the pump momentum move towards
$k_x=2|\mathbf{k}_p|$. Note that close to the lower threshold (c in
Fig. \ref{PumpUnstable})
there are two distinct momenta $k_x$ for the signal and two for the idler
states, where the instability develops.

\onecolumngrid

\twocolumngrid

For a strong pump (high $n_p$), the pump mode becomes stable to small
fluctuations again, defining an `upper threshold'. At the upper threshold the
instability develops at an unique momentum for signal and idler states. The
behaviours above the upper threshold are shown in the bottom row of
Fig. \ref{simultaneouszeros}. Just above the upper threshold, there are only
two places where $\Im(\omega^+(\mathbf{k})) \neq \Im(\omega^-(\mathbf{k}))$:
one near $k_x=0$ and one near $k_x=2|\mathbf{k}_p|$. As the pump strength
is further increased these peaks eventually disappear while the real parts
of the eigenvalues ($\Re(\omega^\pm(\mathbf{k}))$) separate and become
increasingly close to the poles ($\xi^\pm(\mathbf{k})$).

In general, $\xi^\pm(\mathbf{k})$ (solid blue lines in
Fig. \ref{simultaneouszeros}) pinch together at the momenta where
$\Im(\omega^\pm(\mathbf{k}))$ is closest to $0$. The values of
$\xi^\pm(\mathbf{k})$ are very close to $\Re(\omega^\pm(\mathbf{k}))$
apart from where $\Im(\omega^\pm(\mathbf{k}))$ split (or differ from
$\Im(\omega^{\pm}(\mathbf{k}))=-\kappa$). At these points, the effective
chemical potential is equal to the real parts of the eigenvalues. The
phase transition happens where the real and imaginary parts of the
determinant of the inverse retarded Green's function become zero
simultaneously, which indicates diverging luminescence in the normal
state and signals the phase transition. This happens precisely when
$\mu_{\mathrm{eff}}(\mathbf{k})=\xi^\pm(\mathbf{k})$. In this sense the OPO
phase transition happens in an analogous way to an equilibrium BEC phase
transition: the effective chemical potential $\mu_{\mathrm{eff}}(\mathbf{k})$
(red dashed line in Fig. \ref{simultaneouszeros}) moves closer to the
energies of the system defined by $\xi^\pm(\mathbf{k})$ (solid blue lines
in Fi. \ref{simultaneouszeros}) as the density is increased, and the OPO
transition takes place
when $\mu_{\mathrm{eff}}(\mathbf{k})$ reaches $\xi^\pm(\mathbf{k})$.

\begin{figure}[h]
\includegraphics[width=\columnwidth]{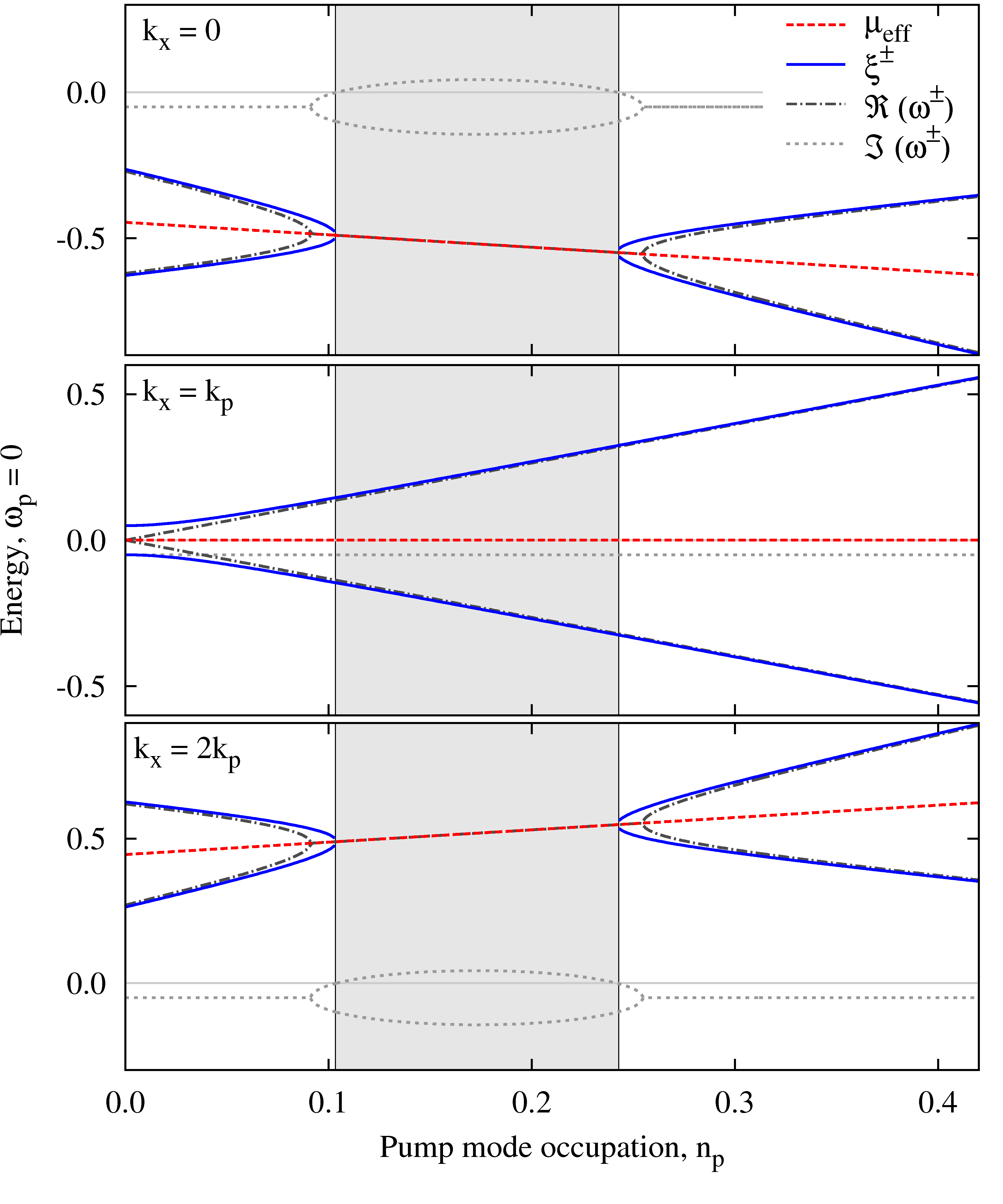}
\caption[]{(Colour online) As Fig. \ref{simultaneouszeros} but at selected
momenta. The unstable region is indicated in grey. Top: $k_x=0$ i.e. at
the expected signal, the chemical potential $\mu_{\mathrm{eff}}$ decreases
with increasing $n_p$ to cross $\xi^\pm$ when $\Im(\omega^{+})=0$; middle:
$k_x=|\mathbf{k}_p|$, $\mu_{\mathrm{eff}}$ never crosses $\xi^\pm$ and
$\Im(\omega^\pm)<0$ for any $n_p$ i.e. there is no instability directly at
the pump; bottom: $k_x=|2\mathbf{k}_p|$ i.e. at the expected idler, the
chemical potential $\mu_{\mathrm{eff}}$ increases with increasing $n_p$
to cross $\xi^\pm$ when $\Im(\omega^{+})=0$. \label{kchoicezeros}}
\end{figure}

To show clearly what happens across the instability threshold, we examine
the behaviour of $\Re(\omega^\pm(\mathbf{k})), \Im(\omega^\pm(\mathbf{k})),
\xi^\pm(\mathbf{k})$ and $\mu_{\mathrm{eff}}(\mathbf{k})$ over a range
of pump mode occupations for specific characteristic momenta. Since
we are interested in the transition to the OPO regime, we choose three
characteristic momenta in the lab frame: $k_x=k_p$, $0$ and $|2\mathbf{k}_p|$
i.e. the locations of the pump and expected signal and idler states. As can
be seen in Fig. \ref{kchoicezeros}, at the pump, $\Im(\omega^\pm)<0$ and
$\mu_{\mathrm{eff}}\ne \xi^\pm$ at any $n_p$ i.e. there is no instability
at the pump momentum at any occupation. This means that with our choice of
the pump being below the LP dispersion we are in the optical limiter and not
bistability regime\cite{PhysRevB.71.115301, PSSB:PSSB200560961}. In contrast,
for $k_x=0$ the effective chemical potential $\mu_{\mathrm{eff}}$ decreases
as the density is increased and crosses $\xi^\pm$ in two places indicating
the upper and the lower threshold. Around the idler, $k_x = 2|\mathbf{k}_p|$,
the effective chemical potential is increasing with increasing density. We
can also see that the mode crossing occurs at the transition from a stable
to an unstable region $\Im(\omega^\pm)=0$. This behaviour is analogous to
other bosonic condensations, where the phase transition is associated with the
chemical potential crossing one of the energy modes.\cite{PhysRevB.75.195331,
BECPitaevskiiStringari} In particular, we note that the increasing chemical
potential and therefore the closest analogy is around the expected idler.

This behaviour of pinching and crossing can be seen clearly from the
expressions for $\mu_{\mathrm{eff}}(\mathbf{k})$ (Eq. \eqref{pumpchempot}),
$\omega^\pm(\mathbf{k})$ (Eq. \eqref{pumpeigens}) and $\xi^\pm(\mathbf{k})$
(Eq. \eqref{pumppoles}). The imaginary parts of $\omega^\pm$ differ
from $-i\kappa_{lp}$ when the discriminant in Eq. \eqref{pumpeigens}
is negative; the real part giving the spectra is then the same as the
chemical potential in Eq. \eqref{pumpchempot}. Meanwhile, the first term in
Eq. \eqref{pumppoles} is also the chemical potential, so when the discriminant
is small the poles will be close to the chemical potential. This happens when
$(\alpha^++\alpha^-)-4V_{p+-}n_p \approx -4\kappa_{lp}^2$, which is exactly
the condition for the imaginary part of one of the eigenvalues to become zero.

\subsection{Eigenvalues of the distribution matrix and effective temperature}

\begin{figure}[h]
\includegraphics[width=\columnwidth]{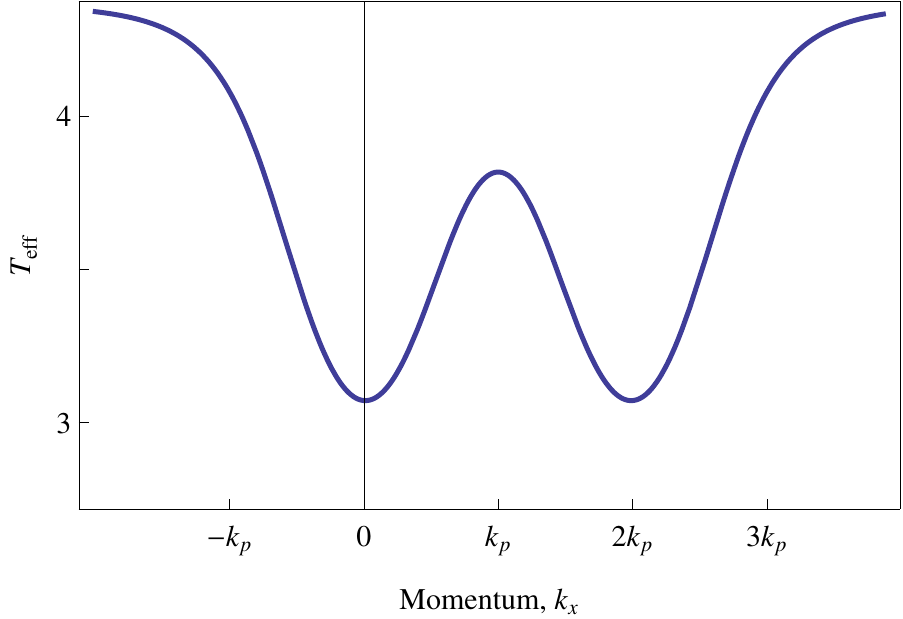}
\caption{(Colour online) The effective temperature,
$T_{\mathrm{eff}}(\mathbf{k})$ as defined by Eq. \eqref{pumpTeff},
with a local maximum at the applied pump and global minima at $k_x = 0,
|2\mathbf{k}_p|$. The pump mode occupation provides a purely multiplicative
factor.\label{Teffplot} }
\end{figure}

The concept of an effective temperature has been introduced in several
driven-dissipative systems. Examples include glassy systems or shaken
sand where there is a separation of time scales of the motion, and the
effective temperature is related to the slow dynamics. \cite{Song15022005,
PhysRevE.80.031132, JPhys.CondMatt.14.1683, PhysRevE.55.3898} More
recently, the extended fluctuation-dissipation relations that appear
in the Keldysh formalism have been used to introduce an effective
temperature for quantum driven-dissipative systems studied using Keldysh
Green's functions. \cite{PhysRevA.87.023831, PhysRevB.89.134310,
ArXiv:1507.01939} In equilibrium, the bosonic distribution matrix,
$F_s$, is given by $\coth(\frac{\omega-\mu}{2T})$ ($k_B =1$), where
$\mu$ is the chemical potential, i.e. $F_s$ diverges as $2T/\omega$
when $\omega\rightarrow \mu$. This relation allows to identify a low
frequency effective temperature, $T_{\rm eff}$, in systems driven away from
equilibrium.\cite{PhysRevA.87.023831} By examining the positive eigenvalue
of $F_s$ ($\lambda_{F_s}^+$), we can define an effective temperature as:
\beq
\lambda_{F_s}^+ =
\frac{2T_{\mathrm{eff}}}{\omega-\mu_{\mathrm{eff}}}. \label{defTeff}
\eeq
The eigenvalues of $F_{s}(\omega,\mathbf{k})$ (Eq. \eqref{pumpFsmatrix}) are:
\beq
\lambda_{F_s}^\pm(\omega,\mathbf{k}) = \pm
\sqrt{1+\frac{4V_{p_{+-}}^2n_p^2}{(2\omega -\alpha^++\alpha^-)^2}}\nonumber
\eeq
and we consider the positive eigenvalue
$\lambda^+_{F_s}(\omega,\mathbf{k})$. For $\omega \sim (\alpha^+-\alpha^-)/2$,
the second term dominates and
\beq
\lambda^+_{F_s}(\omega,\mathbf{k}) \approx
\frac{V_{p_{+-}}n_p}{\omega-\frac{\alpha^+-\alpha^-}{2}}.  \nonumber
\eeq
From Eq. \eqref{pumpchempot}, $(\alpha^+-\alpha^-)/2$ is the effective
chemical potential, $\mu_{\mathrm{eff}}$ and, using Eq. \eqref{defTeff},
the low energy effective temperature
\beq
T_{\rm eff}(\mathbf{k}) = \frac{V_{p_{+-}}n_p}{2}, \label{pumpTeff}
\eeq
which is plotted in Fig. \ref{Teffplot} is obtained. The shape of
$T_{\mathrm{eff}}(\mathbf{k})$ is set by the $X^4$ contribution in
$V_{p_{+-}}$ and has minima at momenta $k_x=0$ and $k_x=|2{\mathbf k}_p|$
in the lab frame. It is interesting to note that in the OPO transition the
``condensation'' happens into signal and idler momenta close to the lowest
effective temperature.

\subsection{Incoherent luminescence, absorption and spectral weight}

When the single mode ansatz is stable, $D^R, D^A$ and $D^K$ are found
using Eqs. \eqref{DRinversion} - \eqref{DKinversion} and the incoherent
luminescence, absorption and spectral weight around the pump-mode
state can be calculated using Eqs. \eqref{dlessgreat}-\eqref{absdef} and
\eqref{swdef}. For the remainder of this section, the pump mode occupations
of Figs. \ref{simultaneouszeros}c and \ref{simultaneouszeros}d, close to
the border of the unstable region, are considered, and all momenta are in
the lab frame.

\begin{figure}[h]
\includegraphics[width=\columnwidth]{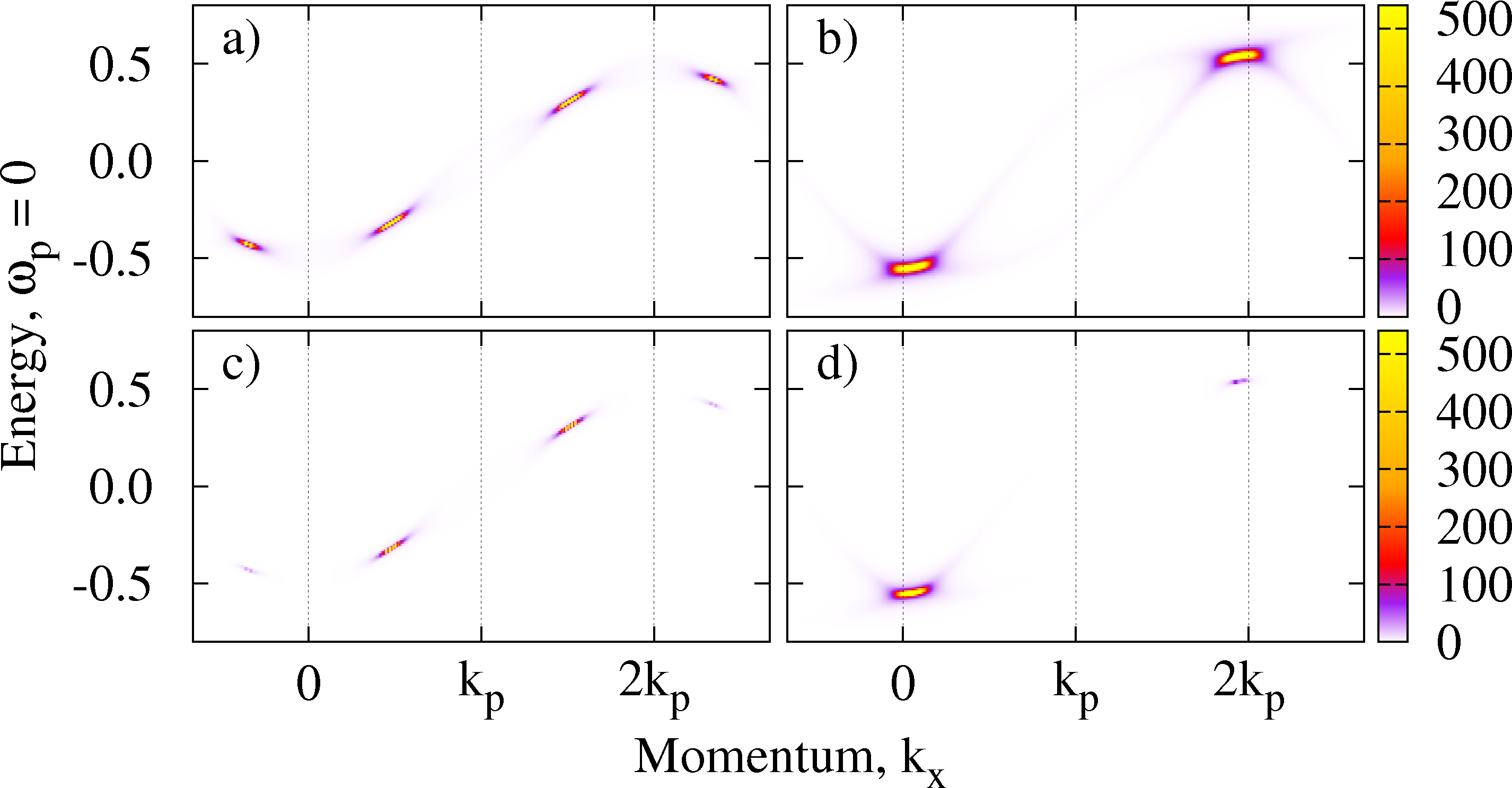}\\
\caption{(Colour online) Incoherent polariton (top) and photon
(bottom) luminescence near the instability thresholds. Left: below
lower threshold $n_p = 0.073$; right: above upper threshold $n_p =
0.245$. \label{PumpIncoherentLum}}
\end{figure}

The incoherent luminescence shows peaks corresponding to the
locations of the splitting in the imaginary parts of the eigenvalues
$\Im(\omega^\pm)$ in Fig. \ref{simultaneouszeros}. When the pump is weak,
below the lower threshold, four peaks appear in around the pump mode,
as shown in Fig. \ref{PumpIncoherentLum}a. Above the `upper threshold',
Fig. \ref{PumpIncoherentLum}b, there are only two peaks centred near $k_x=0$
and $k_x=|2\mathbf{k}_p|$. This is consistent with the behaviour of the
eigenvalues in Figs. \ref{PumpUnstable} and \ref{simultaneouszeros}.

One effect of assuming the same polariton decay rate at all momenta is to
make the intensity peaks in the polariton luminescence appear symmetric
about the pump mode which reflects the pairwise scattering process. In
experiments, only the photonic component of polaritons can be measured
and so the signal, which is more photon-like, appears stronger than the
idler. \cite{PhysRevB.62.16247, PhysRevB.70.205301} Thus, in the lower panels
of Fig. \ref{PumpIncoherentLum}, the luminescence is rescaled according to
the photon fraction (Eq. \eqref{PhotRescale}) and the pure photon luminescence
is stronger at low momenta, as expected.

\begin{figure}[h]
\includegraphics[width=\columnwidth]{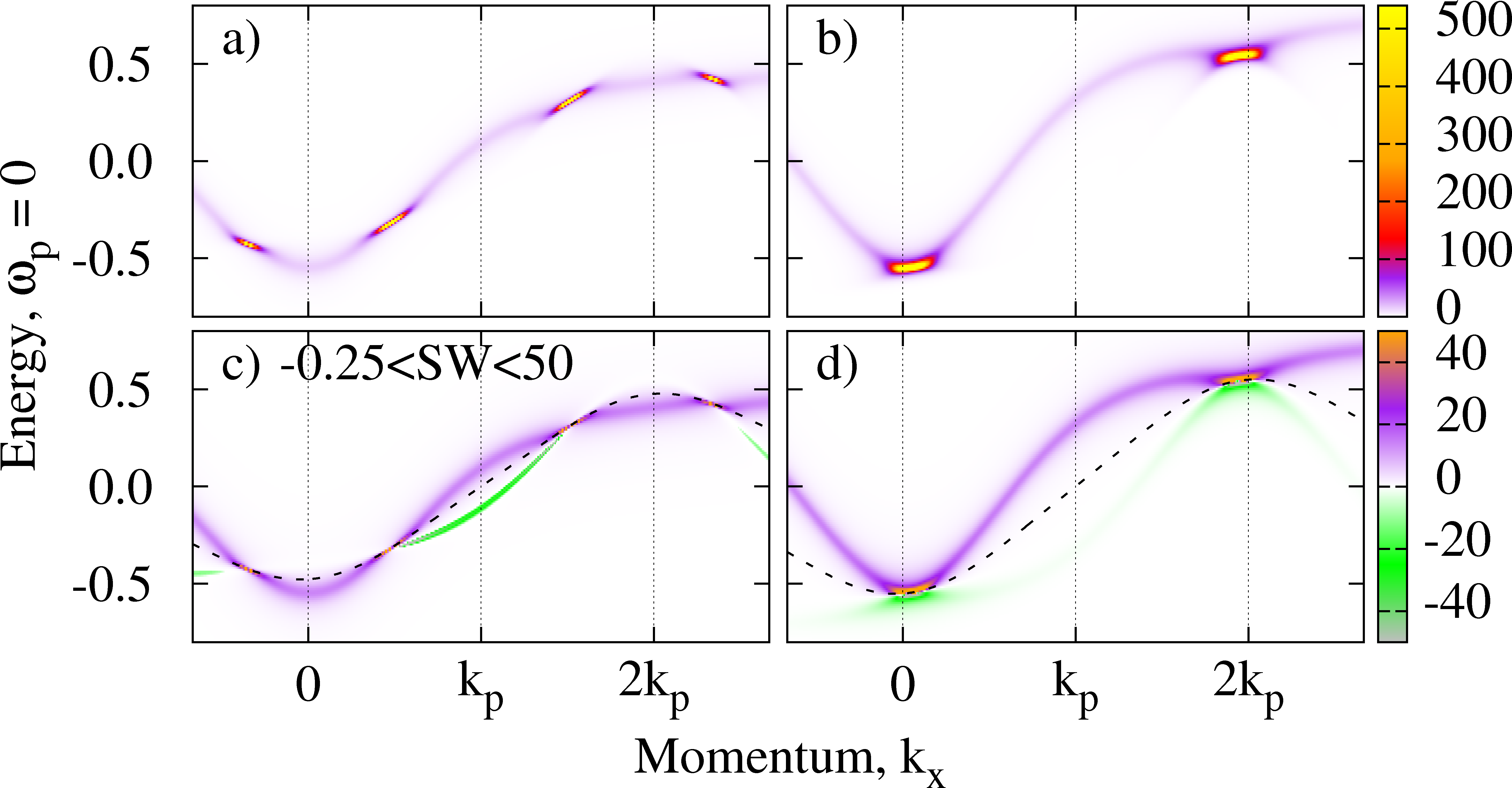}\\
\caption{(Colour online) Polariton absorption (upper) and spectral
weight (lower) near the instability thresholds. Left: below
lower threshold $n_p = 0.073$; right: above upper threshold $n_p =
0.245$. The scales for positive spectral weight are the same, at weak
pumping, there is only weak negative spectral weight and the range
is curtailed for visibility. The dashed line in the lower panels is
the effective chemical potential $\mu_{\mathrm{eff}}(\mathbf{k})$ of
Fig. \ref{simultaneouszeros}. \label{PumpAbsSW}}
\end{figure}
The absorption follows the same general pattern as the luminescence, but is
generally stronger on the upper branch of the spectrum.

It is also interesting to examine the spectral weight (Eq. \eqref{swdef}). As
seen in Figs. \ref{PumpAbsSW}c and \ref{PumpAbsSW}d, there are regions
of negative spectral weight where the luminescence is greater than the
absorption. Above the upper threshold, the spectral weight is negative
below the chemical potential, as it is usually the case. At weak pumping,
the spectral weight is only very weakly negative, this occurs for energies
below the chemical potential and away from the peaks in the luminescence. In
Fig. \ref{PumpAbsSW}c, the negative part of the spectral weight range is
greatly reduced to show the negative spectral weight.

Finally, the luminescence is integrated over energy and plotted as a function
of two-dimensional momentum with ${\bf k} = (k_x,k_y)$. What looked as
four peaks in the ($\omega$, $k_x$) plots (Fig. \ref{PumpIncoherentLum}
and \ref{PumpAbsSW}) was a signature of a ring structure for the signal and
idler. At low pump powers close to the lower threshold the pump-only state
becomes unstable to a signal state with a ring shape in momentum. For large
pump occupations, just above the `upper threshold', the instability develops
at a unique momentum ${\bf k}_s=(k_s,0)$ and so only two distinct peaks
associated with developing signal and idler states are observed. However,
there is no distinct signal momentum when the pump mode first becomes
unstable, but there is instead initial growth at a range of momenta on a
ring $|{\bf k}_s-(\Delta k_x,0)|=k_s-\Delta k_x$ before a single momentum
value dominates. The OPO transition can therefore be described using a
distinct \textit{pair} of new modes if the pump is decreased through the
`upper threshold' of the instability.

\begin{figure}[h]
\includegraphics[width=\columnwidth]{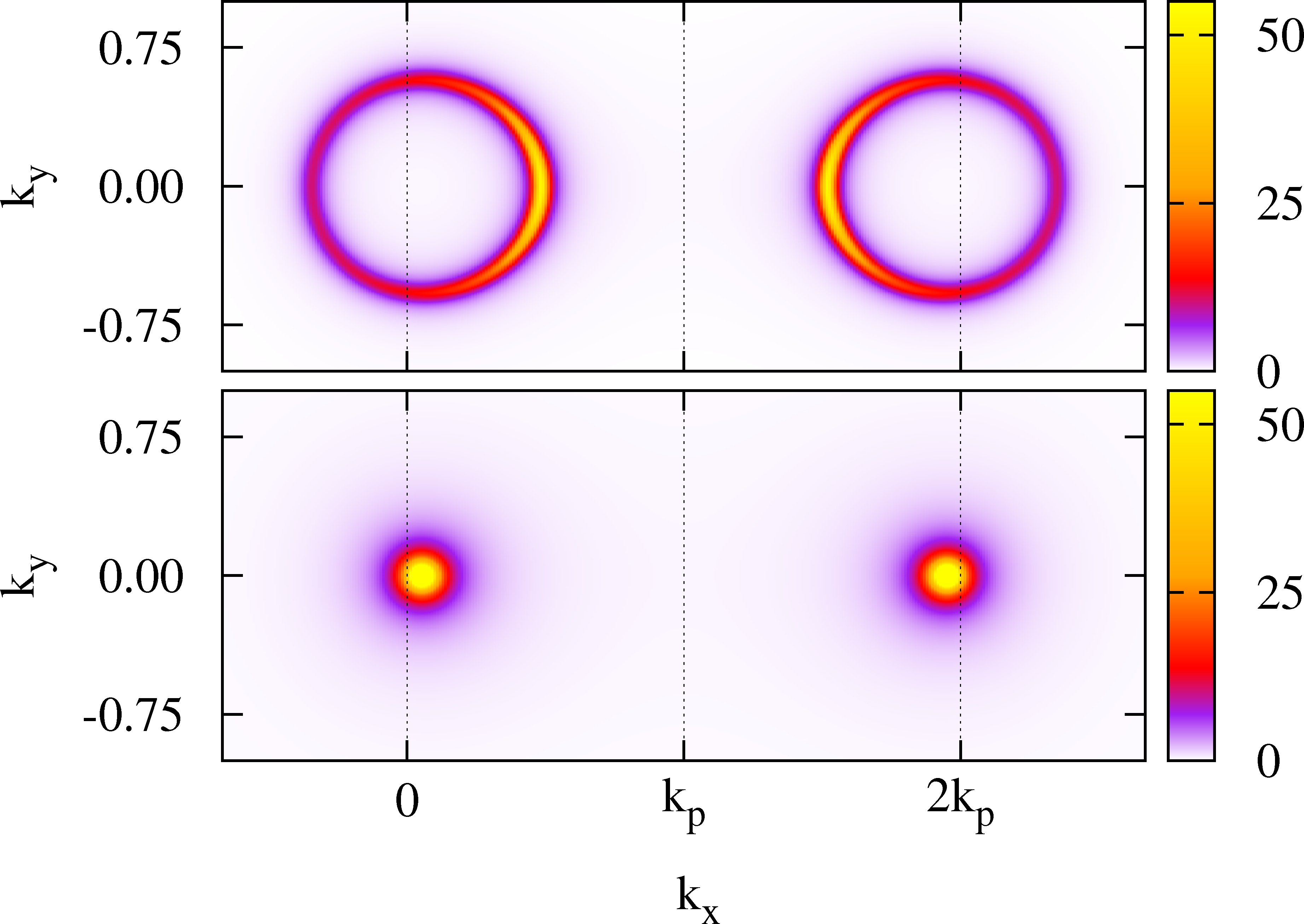}
\caption{(Colour online) Incoherent polariton luminescence integrated over
energy and plotted in 2-D momentum space. Top: $n_p=0.073$; bottom: $n_p
=0.245$. \label{kxkyluminescence}}
\end{figure}

\begin{figure}[h]
\includegraphics[width=\columnwidth]{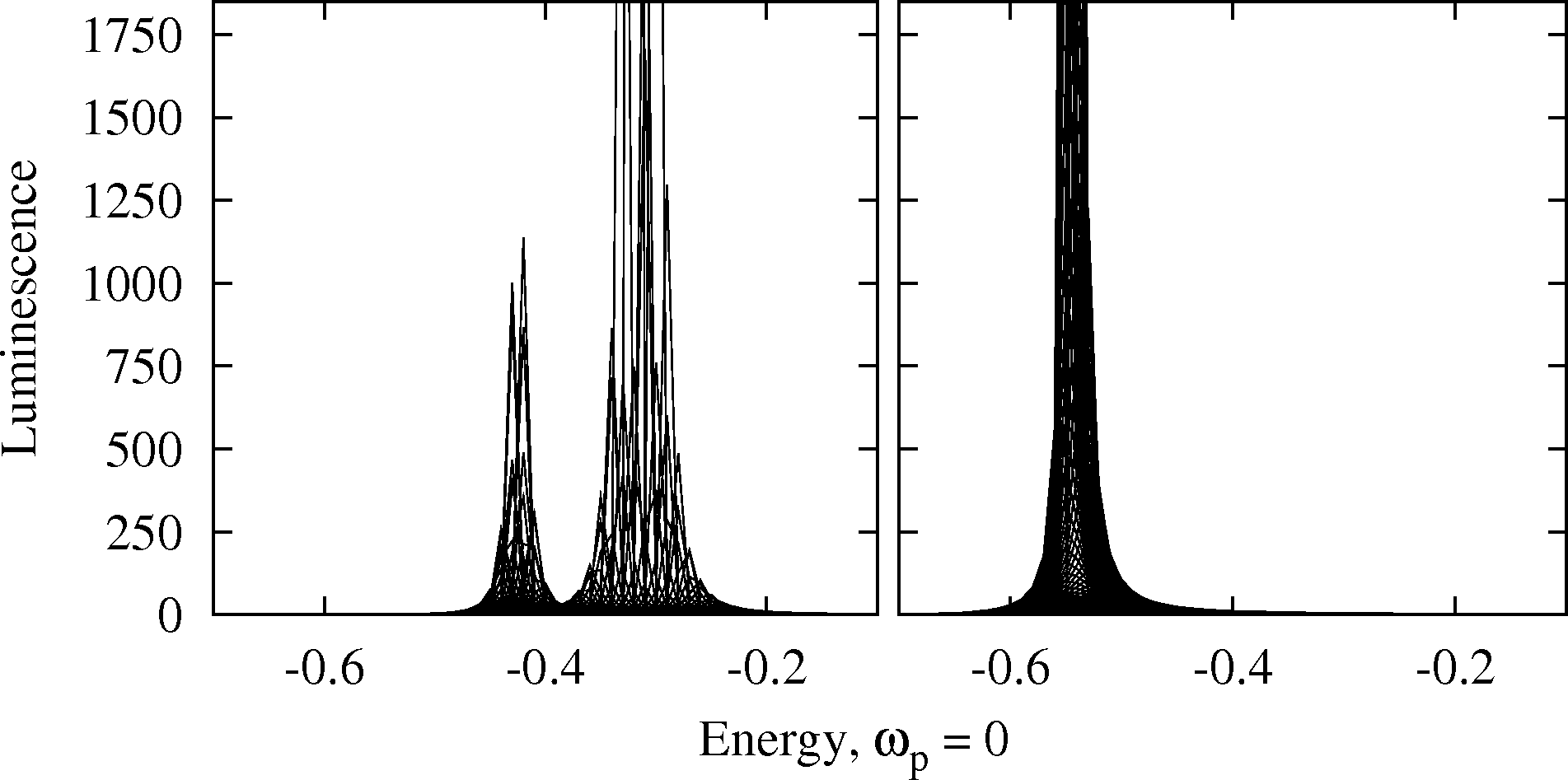}
\caption{Incoherent luminescence at energies below $\omega_p$ for a large
rang of momenta. Left: $n_p=0.073$, below the lower threshold, the peak
at higher energy dominates; right: $n_p=0.245$ above the upper threshold,
the single peak is at a lower energy and is much narrower than the two peaks
present for a weaker pump. \label{PumpLumkSlices}}
\end{figure}

In Fig. \ref{PumpLumkSlices}, the incoherent luminescence is plotted for
all momenta ${\bf k}=(k_x, 0)$ and energies below the pump energy (around
the signal only). For weak pumping, the two peaks are broad and the one at
the lower energy is weaker than the one at higher energy. Comparing this to
Fig. \ref{PumpIncoherentLum}a, this shows that the peak at higher momentum
dominates, which is consistent with the weaker luminescence on the side of
the ring away from the pump in Fig. \ref{kxkyluminescence}. Above the `upper
threshold', the peak in the luminescence is narrower in energy signalling
a phase transition to a single pair of signal-idler modes.

\section{Above OPO threshold}\label{OPOcalculations}

In the previous section the precursor of the OPO transition which manifested
itself by the appearance of large occupations near ${\mathbf k}=0$ and
${\mathbf k}=2{\mathbf k}_p$, and had a particularly simple structure above
the `upper threshold' was examined. Here the analysis is extended to the
regime where the pump-only state mean-field solution discussed before was
unstable to small fluctuations. The next level of complexity is considered
by including two additional modes in the mean-field ansatz.

\subsection{OPO states and action}

To examine the behaviour for parameters where the pump-only mean field
solution is unstable, the field is divided into three subspaces in energy and
momenta around the signal, pump and idler states such that $\psi = \psi_s +
\psi_p +\psi_i.$ The general idea is to include two additional modes, the
signal, $s$, and idler, $i$, into the mean field ansatz, where each mode is
restricted to momenta $\mathbf{q}_j +\mathbf{k}_p = \mathbf{k}_s, \mathbf{k}_p,
\mathbf{k}_i$,\cite{PhysRevA.76.043807, PhysRevB.71.115301, PhysRevB.75.075332}
and then look at fluctuations around this solution. Due to the complicated
nature of the instabilities when the pump strength is increased towards the
lower threshold (low $n_p$), this simple mean field ansatz is valid in the
region of stronger pump occupations closer to the upper threshold, where the
OPO transition is expected to occur in a straightforward manner, i.e. where
the peaks in the luminescence are located at two distinct momenta only with
$k_y = 0$, as shown in the lower panel of Fig. \ref{kxkyluminescence}.

The full Keldysh action, Eq. \eqref{PartsofAction}, after integrating out
the polariton decay bath, where the fields have been formally divided into
three sub-spaces around signal, pump and idler is:
\beqy
&& S_{OPO} = \int d t \Big(-\sqrt{2}f(\bar{\psi}_{p,q}+\psi_{p,q}) +
\nonumber \\
&& \sum_{j=s,p,i}\Big[\bar{\psi}_{j,cl}(i\partial_t-\omega_{lp}({\mathbf
k}+{\mathbf k}_p)+\omega_p-i\kappa_{lp}) \psi_{j,q} + \nonumber \\
&& \bar{\psi}_{j,q}(i\partial_t-\omega_{lp}({\mathbf k}+{\mathbf k}_p)+\omega_p
+i\kappa_{lp})\psi_{j,cl} )\Big] - \nonumber \\
&&
\Big\{\sum_{j=s,p,i}\frac{V_{jjjj}}{2}(\bar{\psi}_{j,cl}\bar{\psi}_{j,q}(\psi_{j,cl}^2
+\psi_{j,q}^2) + \nonumber \\
&& V_{sisi}\big[(\bar{\psi}_{s,cl}\bar{\psi}_{i,cl}+
\bar{\psi}_{s,q}\bar{\psi}_{i,q})(\psi_{s,cl}\psi_{i,q}+\psi_{s,q}\psi_{i,cl})
\big] + \nonumber \\
&& \sum_{j=s,i} V_{pjpj}\big[(\bar{\psi}_{p,cl}\bar{\psi}_{j,cl}+
\bar{\psi}_{p,q}\bar{\psi}_{j,q})(\psi_{p,cl}\psi_{j,q}+\psi_{p,q}\psi_{j,cl})
\big] \nonumber \\
&& + \frac{V_{ppsi}}{2}\big[2(\bar{\psi}_{s,cl}\bar{\psi}_{i,cl}+
\bar{\psi}_{s,q}\bar{\psi}_{i,q})\psi_{p,cl}\psi_{p,q} + \nonumber \\
&& (\bar{\psi}_{s,cl}\bar{\psi}_{i,q}+
\bar{\psi}_{s,q}\bar{\psi}_{i,cl})(\psi_{p,cl}^2+\psi_{p,q}^2) \big] +
h.c. \Big\} \Big) + \nonumber \\
&& \sum_{j=s,p,i}2i\kappa_{lp}\iint dt dt^\prime
\bar{\psi}_{j,q}(t)F_\chi(t-t^\prime)\psi_{j,q}(t^\prime) \label{OPOaction}
\eeqy
where the fields for each mode $m$ are implicitly of the form
$\psi_{m,\{cl,q\}}(t)$. The interaction coefficients now have 4 indices that
indicate exactly which modes are involved in each scattering process.

\subsection{Mean Field}

Taking the functional derivatives with respect to all $\psi$ fields in
Eq. \eqref{OPOaction} and setting them to zero leads to the set of mean-field
equations analogous to Eq. \eqref{generalcGPE}. A mean field ansatz formed of
three plane waves is chosen. \cite{PhysRevB.71.115301, PhysRevB.75.075332,
PhysRevA.76.043807} Written relative to the pump as before, the non-zero
classical saddle-point fields are:
\beqy
\psi^{\mathrm{sp}}_{s,cl}(t,{\mathbf x}) &=& \sqrt{2}
Se^{i\tilde{\omega}t}e^{-i\tilde{\mathbf k}\cdot \mathbf x}, \nonumber \\
\psi^{\mathrm{sp}}_{p,cl}(t,{\mathbf x}) &=& \sqrt{2} P, \nonumber \\
\psi^{\mathrm{sp}}_{i,cl}(t,{\mathbf x}) &=& \sqrt{2} I
e^{-i\tilde{\omega}t}e^{i\tilde{\mathbf k}\cdot \mathbf x}, \nonumber
\eeqy
with the signal and idler energies $\omega_{s,i} = \omega_p \mp \tilde{\omega}$
and momenta $\mathbf{k}_{s,i} = \mathbf k_p \mp \tilde{\mathbf k}$. The
mode amplitudes $S, P, I$ are the mean-field amplitudes and so have the
pre-factor $\sqrt{2}$ as discussed in section \ref{MFSPsbackground}. The
general saddle-point form of any mode in the quantum-classical basis is:
\beq
\psi^{\mathrm{sp}}_{m,cl}(t,{\mathbf x}) = \sqrt{2} M
e^{-i\omega_mt}e^{i\mathbf q_m \cdot \mathbf x}; \quad \psi^{\mathrm{sp}}_{m,q}
= 0 \label{OPOmodeform}
\eeq
where, given the gauge transformation to the pump frame, the momenta
are: $\mathbf{q}_s = -\tilde{\mathbf{k}}$, $\mathbf{q}_p = \mathbf{0}$,
$\mathbf{q}_i = \tilde{\mathbf{k}}$ and energies: $\omega_s = -\tilde{\omega}$,
$\omega_p = 0$ and $\omega_i = \tilde{\omega}$ in what follows.

We could also substitute the new mean-field ansatz, where the three modes
of interest defined in Eq. \eqref{OPOmodeform} are included explicitly,
\beq
\psi^{\mathrm{mf}}(t,\mathbf{x}) =
Se^{i\tilde{\omega}t}e^{-i\tilde{\mathbf{k}}\cdot
\mathbf{x}}+P+Ie^{-i\tilde{\omega}t}e^{i\tilde{\mathbf{k}}\cdot \mathbf{x}}
\nonumber
\eeq
directly into the general cGPE, Eq. \eqref{generalcGPE}, and take the
steady state with \mbox{$\partial_t P = \partial_t S = \partial_t I =
0$}. This gives three complex equations that can be solved to give the
signal energy $\omega_s=-\tilde{\omega}$ and the complex mode amplitudes $S,
P, I$. \cite{PhysRevB.71.115301} Some of the interaction terms introduce
modes outside of the three mode ansatz; these are discarded. The cGPEs
for each of the modes, after substitution of the three mode ansatz
are\cite{PhysRevB.71.115301}
\beqy
{\Xi_s S + V_{sppi}P^2I^*} &=&0 , \label{OPOsignal}\\
{\Xi_p P + 2V_{sppi}SP^*I +f} &=&0, \label{OPOpump} \\
{\Xi_i I + V_{sppi}S^*P^2} &=&0, \label{OPOidler}
\eeqy
where the shorthand
\beqy
\Xi_m &=& \omega_{lp}(\mathbf{q}_m+\mathbf{k}_p)+
2(V_{mmss}n_s+V_{mmpp}n_p+V_{mmii}n_i)\nonumber \\
&& - \omega_m-\omega_p -V_{mmmm}n_m -i\kappa_m, \nonumber
\eeqy
and $|P|=n_p$, $|S|=n_s$, $|I|=n_i$ has been used. Since the polariton
decay is constant, $\kappa_s = \kappa_i = \kappa_p = \kappa_{lp}$. The
steady state requires that the signal and idler momenta are specified;
\cite{PhysRevB.75.075332} we work with the simplest choice of ${\mathbf k}_s
= {\mathbf 0}$ so ${\mathbf k}_i = 2{\mathbf k}_p$, although in experiments
${\mathbf k}_s$ is usually small but finite. \cite{PhysRevB.68.115325,
PhysRevB.92.035307} In the OPO regime, the occupation of the pump mode is
depleted due to scattering into the signal and idler modes. The complex mode
amplitudes can be considered to have the form $M = |M|e^{i\theta_m}$. The
phase of the pump mode is locked to the external pump, and can be determined
from Eq. \eqref{OPOpump} but there is a phase freedom in the choice of
the signal and idler, their phase difference is free. It is spontaneously
chosen at each realisation of an experiment. \cite{PhysRevA.76.043807} In the
calculation of the mean field, we are thus free to choose the phase of one of
these remaining modes. We choose the signal to be real and the idler phase
is then determined by the steady-state equations (Eqs. \eqref{OPOsignal} -
\eqref{OPOidler}). This phase freedom means that a shift of the signal phase
e.g. $\theta_s \rightarrow \theta_s + \Delta\theta$, would be accompanied by
a simultaneous change in the idler phase in the opposite direction, $\theta_i
\rightarrow \theta_i - \Delta\theta$, while the equations of motion remain
unchanged.\cite{PhysRevA.76.043807} This phase freedom leads to the appearance
of a gapless Goldstone mode that is not present in the pump-only configuration.

\begin{figure}[h]
\includegraphics[width=\columnwidth]{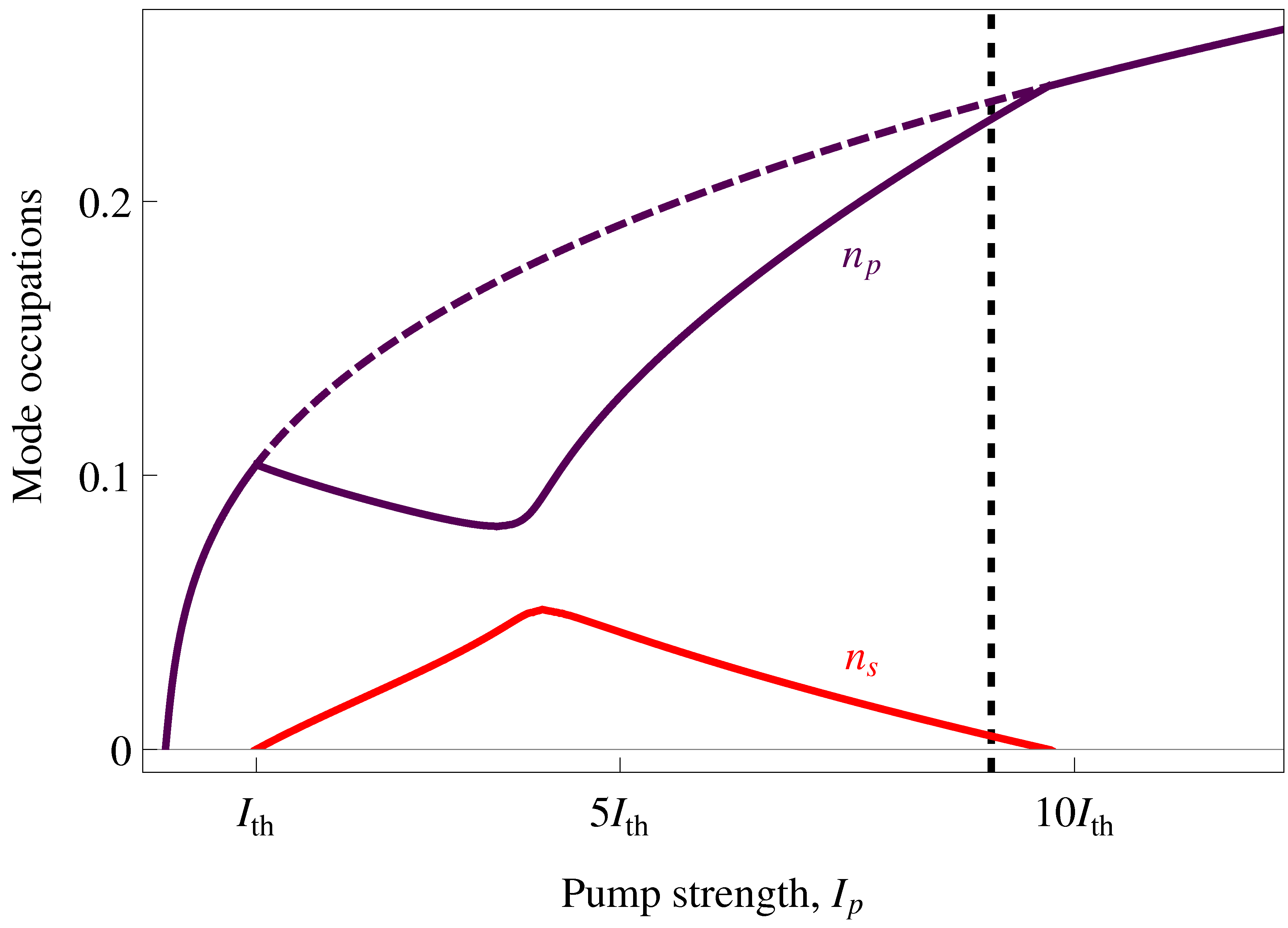}\\
\caption{(Colour online) Signal, $n_s$ (red), and pump, $n_p$ (purple,
the dashed part is the pump only ansatz within the OPO region) mode
occupations within the OPO regime for ${\mathbf k}_p =(1.5,0)$ and
${\mathbf k}_s = (0,0)$. The dashed vertical line is the pump strength
considered for Figs. \ref{StableOPOGoldstone}-\ref{OPOfullLum}. The idler
occupation $n_i$ is the same as the signal occupation for constant polariton
decay.\cite{SemicondSciTech.18.279, PhysRevB.71.115301, PhysRevB.63.193305}
\label{OPOmf}}
\end{figure}

\subsection{Inverse Green's Functions}

The inverse Green's functions for the OPO state are calculated as in the pump
only case ($\psi_{m,cl} \rightarrow \psi^{sp}_{m,cl} + \delta \psi^m_{cl},
\psi_{m,q} \rightarrow \delta\psi^m_{q}$), and give $6\times 6$ matrices due
to the presence of three modes. Using the form of the fluctuations defined
in Eqs. \eqref{clfluctuations} and \eqref{qfluctuations}, the mean-field
plus fluctuations in the Keldysh quantum-classical basis is:
\beqy
\psi_{\mathbf{k},cl}(\omega) &=&
\sqrt{2}S\delta_{\mathbf{k},-\tilde{\mathbf{k}}}\delta_{\omega,-\tilde{\omega}}
+ \delta\psi^s_{\mathbf{k}-\tilde{\mathbf{k}}, cl}(\omega-\tilde{\omega})
\nonumber \\
& & + \sqrt{2}P\delta_{\mathbf{k},\mathbf{0}}\delta_{\omega,0} +
\delta\psi^p_{\mathbf{k}, cl}(\omega) \nonumber \\
& & + \sqrt{2}I\delta_{\mathbf{k},\tilde{\mathbf{k}}}
\delta_{\omega,\tilde{\omega}} + \delta\psi^i_{\mathbf{k}+\tilde{\mathbf{k}},
cl}(\omega+\tilde{\omega}) \nonumber
\eeqy
and
\beq
\psi_{\mathbf{k},q}(\omega) = \delta
\psi^s_{\mathbf{k}-\tilde{\mathbf{k}},q}(\omega-\tilde{\omega})
+ \delta\psi^p_{\mathbf{k},q}(\omega) + \delta
\psi^i_{\mathbf{k}+\tilde{\mathbf{k}},q}(\omega+\tilde{\omega}), \nonumber
\eeq
where the energies and momenta of the modes appear as offsets to the
fluctuations. The inverse Green's functions can be written compactly
as:\cite{PhysRevA.76.043807}
\beq
[D^{-1}]^R(\omega, \mathbf{k}) =\frac{1}{2}\bpm M(+) & Q(+) \\ Q^*(-) &
M^*(-) \epm \label{Dm1ROPO}
\eeq
with the elements of the sub-matrices $M, Q$:
\beqy
M_{m,n}(\pm) &=& \delta_{m,n}(\omega_m \pm\omega-\omega_{lp}({\mathbf k}_m^\pm)
+ \omega_p +i\kappa_{lp}) \nonumber \\
&& \; -2 \sum_{r,t=1}^3\delta_{m+r,n+t}V_{m^\pm,n^\pm,r,t}\psi_{r}^{\rm{mf}
*}\psi_{t}^{\rm{mf}}, \nonumber \\
Q_{m,n}(\pm) &=&
-\sum_{r,t=1}^3\delta_{m+n,r+t}V_{m^\pm,n^\pm,r,t}\psi_{r}^{\rm
{mf}}\psi_{t}^{\rm{mf}}, \nonumber
\eeqy
where $m^\pm = {\mathbf k}_{m}^\pm = {\mathbf q}_m+\mathbf{k}_p\pm \mathbf{k}$,
$\psi^{\rm{mf}}_{m} = \psi_{m,cl}/\sqrt{2}$ and $m,n,r,t \in \{1,2,3\}
\rightarrow \{S,P,I\}$. $[D^{-1}]^R(\omega, \mathbf{k})$ is related to
the linear response matrix as in the pump only case. The inverse Keldysh
Green's function, $[D^{-1}]^K(\omega, \mathbf{k})$, is similar to the one
for pump-only state (Eq. \eqref{PumpDm1K}) with diagonal elements:
\beq
K_{m,n}(\pm) = i\kappa_{lp}F_\chi(\omega_{m}+\omega_p\pm\omega)\delta_{m,n},
\nonumber
\eeq
and
\beq
[D^{-1}]^K(\omega, \mathbf{k}) = \bpm K(+) & 0 \\ 0 & K(-) \epm.  \nonumber
\eeq

\begin{figure}[h]
\includegraphics[width=\columnwidth]{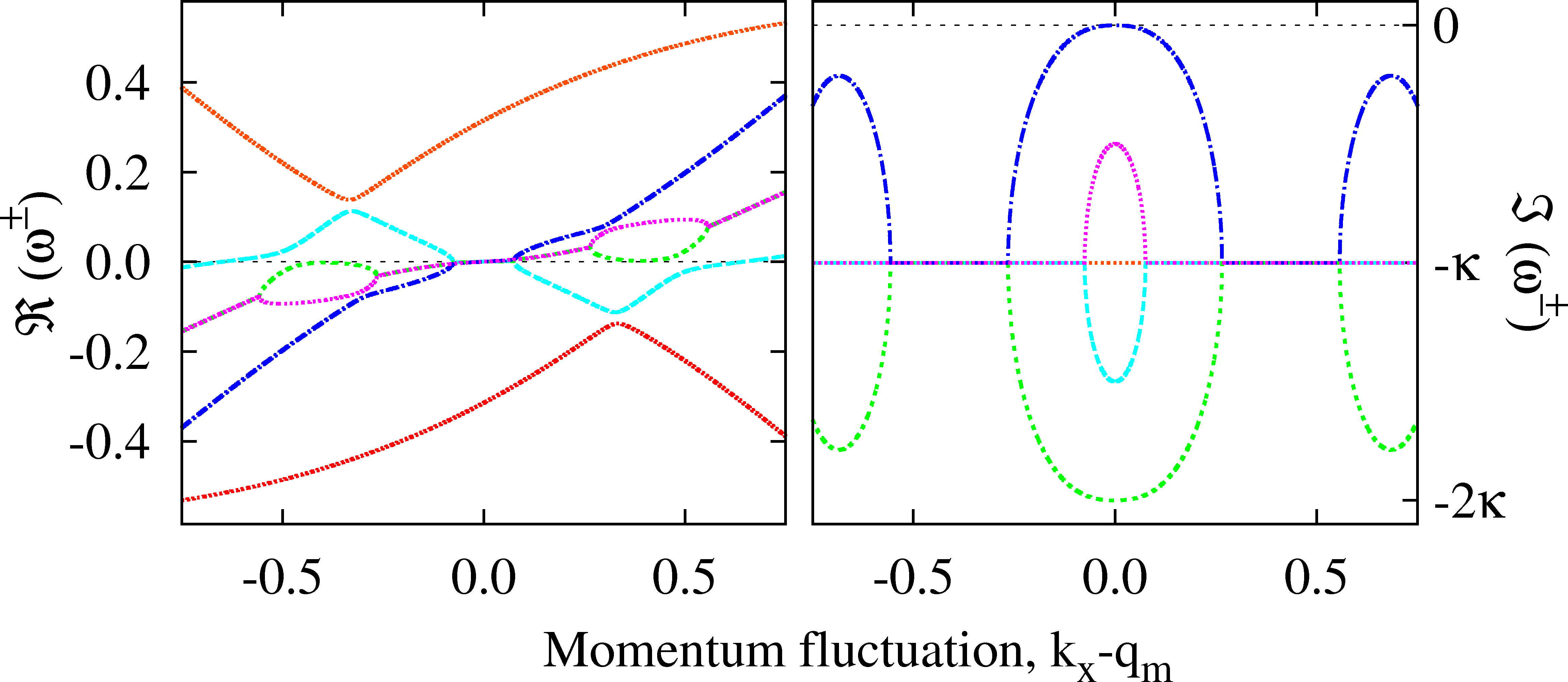}
\caption{(Colour online) Real $\Re(\omega^j)$ and imaginary parts
$\Im(\omega^j)$ of the eigenvalues for the OPO state at pump strength: $I_p =
9.016I_{th}$. The dark blue curve corresponds to the Goldstone mode. Since
all the imaginary parts of the poles are negative, the OPO ansatz is
stable. \label{StableOPOGoldstone}}
\end{figure}

Taking the determinant of Eq. \eqref{Dm1ROPO}, and solving
$\det([D^{-1}]^R(\omega,{\mathbf k}))=0$ for $\omega^j \in \C$ gives the modes
of the system $\omega^j$. With the three signal, idler and pump mean-field
modes there are now six poles. We consider the example of a stable OPO near
the upper threshold as identified in Fig. \ref{OPOmf} ($I_p =9.016I_{th}$). The
real and imaginary parts of the eigenvalues of Eq. \eqref{Dm1ROPO} are plotted
in Fig. \ref{StableOPOGoldstone} showing that the steady state is stable
($\Im(\omega)<0$), and that the Goldstone mode, characterised by $\Im(\omega)
\rightarrow 0 $ {\it and} $\Re(\omega)\rightarrow 0 $ for $k \rightarrow 0$,
is present. \cite{PhysRevA.76.043807} The Goldstone mode is associated with
the spontaneous symmetry breaking of the phase freedom of the difference in
signal and idler phases in the OPO regime. To remain within the three mode
ansatz, it is assumed that the fluctuations in each mode are close in momenta
and energy to that mode and the momentum range for plotting the results is
therefore restricted to $k_x-q_m = (k_p-k_s)/2$.

\begin{figure}[h]
\includegraphics[width=0.85\columnwidth]{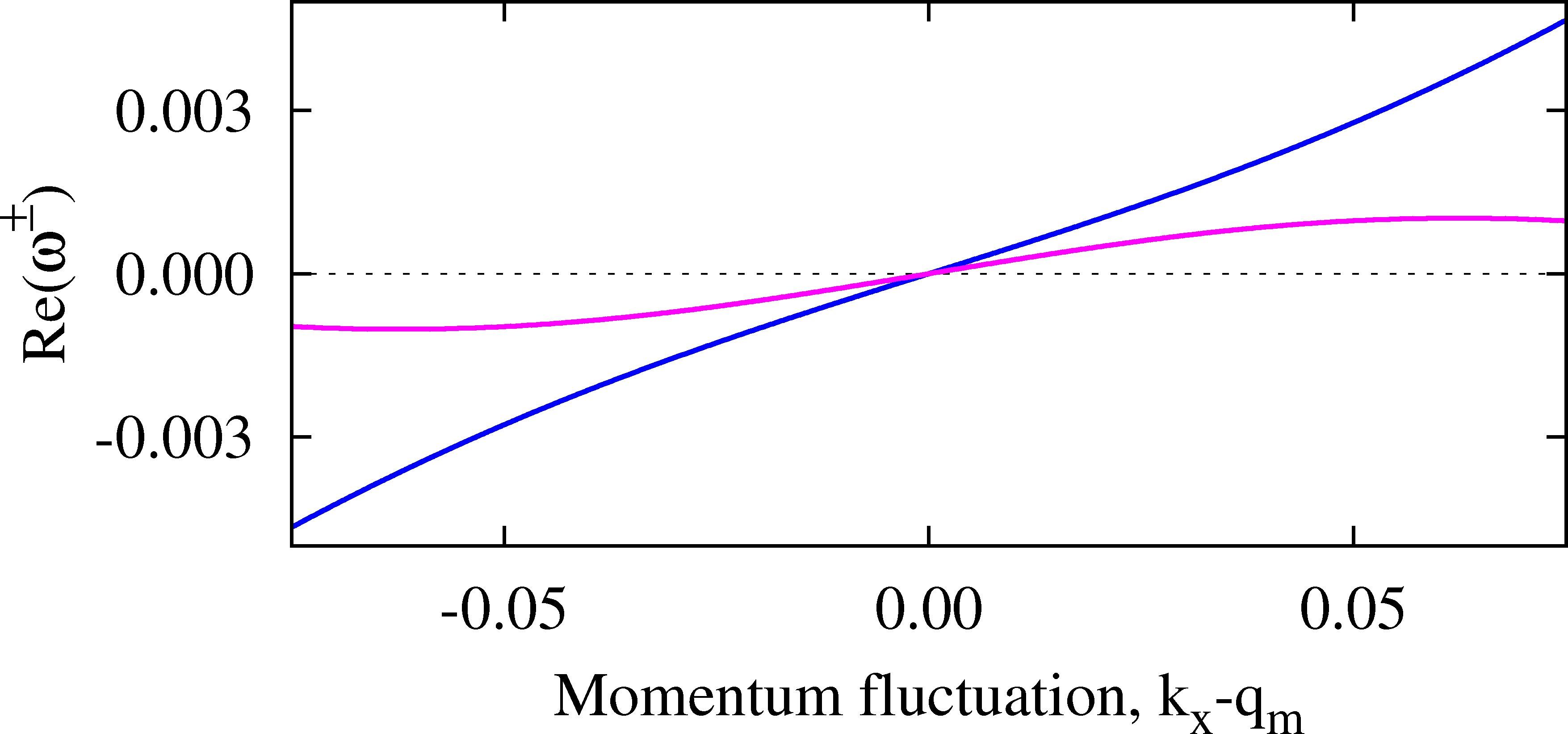}
\caption{(Colour online) The real parts of the spectra in a very small
region around $\delta k = 0$ showing that, although $\mathbf{k}_s = (0,0)$,
there is still a finite slope of the Goldstone mode. \label{spectradetail}}
\end{figure}

In Fig. \ref{spectradetail}, the very central region of the spectra is
plotted. Although in Fig. \ref{StableOPOGoldstone} the real parts of
the spectra appear flat in the limit $\omega \rightarrow 0, \mathbf{k}
\rightarrow 0$, in Fig. \ref{spectradetail}, it is clear that although
$\mathbf{k}_s=(0,0)$, the spectra are still sloped.

\subsection{Luminescence around the OPO states}

\begin{center}
\begin{figure*}[h]
\includegraphics[width=\textwidth]{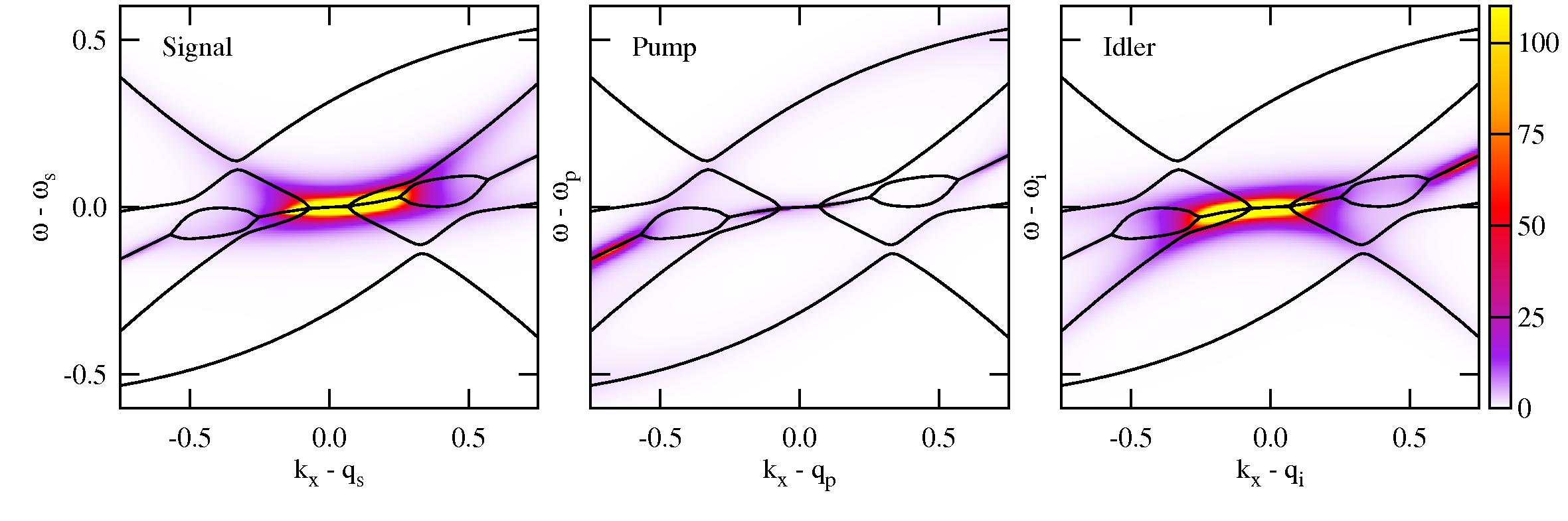}
\caption{(Colour online) Incoherent polariton luminescence about the three
OPO states with the spectra ($\Re(\omega)$) overlaid. The Goldstone mode
dominates the signal and idler states, where the luminescence is strong near
the modes ($\omega_{s,i},k_{s,i}$), while the incoherent luminescence around
the pump is much weaker. \label{OPOmodesLum}}
\end{figure*}
\end{center}

For the stable OPO state, the incoherent luminescence coming from
fluctuations around the three mode ansatz is calculated using
Eqs. \eqref{DRinversion}-\eqref{lumdef}. In Fig. \ref{OPOmodesLum},
the polariton luminescence around the signal, pump and idler modes is
considered separately and the spectra ($\Re(\omega)$ from linear response)
overlaid. Variations in the occupations of the different branches according to
the mode considered become clearly visible. For example, the outermost branches
with increasing energy as the momentum of fluctuations increase are only
noticeably occupied around the pump mode, while the parts of these branches
characterised by decreasing energy with increasing momentum contribute to the
luminescence around the signal mode for negative momentum of fluctuations and
around the idler for positive momentum. The divergence caused by the Goldstone
mode at $\omega_{s,i}, k_x = 0$ leads to significant peaks close to the signal
and idler states. There is only a weak peak in the incoherent luminescence
around the pump mode ($\omega_p, k_x =0$), which is due to the secondary
splitting in the imaginary parts of the eigenvalues (central (blue/green)
lines in the right hand side of Fig. \ref{StableOPOGoldstone}): in this case
since the imaginary part is not zero, the luminescence does not diverge.

In the `normal state', the peaks in the luminescence
(Fig. \ref{PumpIncoherentLum}) coincided with the maximum values of the
imaginary parts of the eigenvalues (Fig. \ref{simultaneouszeros}). In the
`condensed state' the strongest peaks in the incoherent luminescence about
the OPO states are associated with the Goldstone mode while weaker peaks
are the features of where other pairs of the six $\Im(\omega)$ split.

\onecolumngrid
\twocolumngrid

\begin{figure}[h]
\includegraphics[width=\columnwidth]{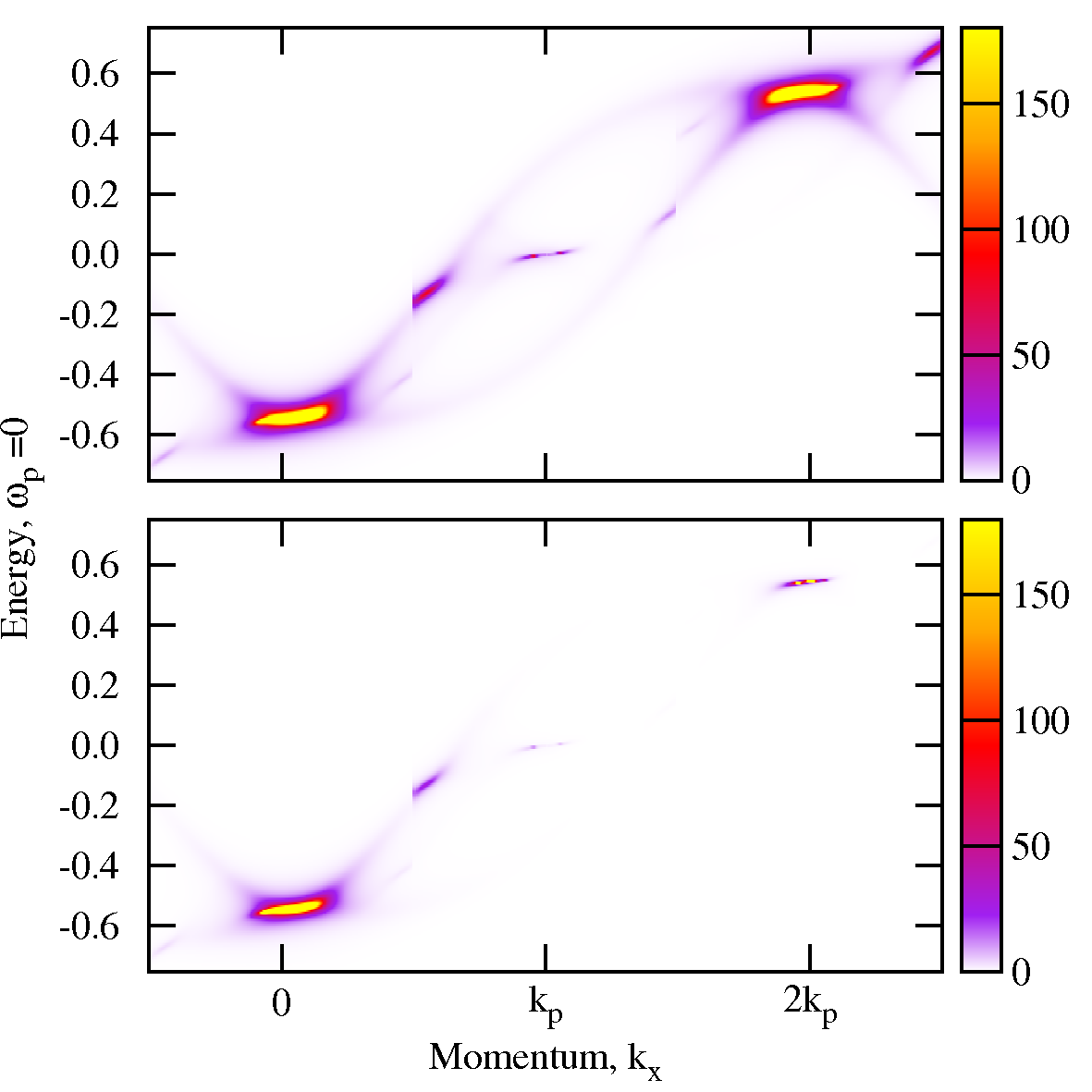}
\caption{(Colour online) The incoherent luminescence around the three OPO
states combined. Top: polariton, Bottom: photon.\label{OPOfullLum}}
\end{figure}

In Fig. \ref{OPOfullLum} the energy and momentum ranges of
Fig. \ref{OPOmodesLum} are shifted to the modes ($ k_x = 0 \rightarrow q_m$
and $\omega_m = 0 \rightarrow \omega_p = 0$) to create a full picture of the
incoherent luminescence around the OPO. The photon parts are included for
completeness, and to highlight the difference in visibility around the three
modes due to the rate at which photons escape. \cite{PhysRevB.70.205301}
In particular, the weak peak at the pump mode becomes insignificant, and
the incoherent luminescence is concentrated around the signal with a very
small region around the idler mode, which are both due to the Goldstone mode.

\section{Conclusions}\label{SummaryConc}

In this article, we have developed a Keldysh functional integral formulation
for a coherently pumped polariton system coupled to a single incoherent
decay bath and have studied the polariton OPO transition. We first examined
the ``normal state'' of the pump only mode and calculated the spectra and
instabilities, in agreement with previous studies. \cite{PhysRevB.75.075332,
PSSB:PSSB200560961, PhysRevB.71.115301} An effective chemical potential
associated with the divergence of the system's bosonic distribution
function, in analogy with the equilibrium Bose-Einstein distribution, was
found. Considering how the ``normal state'' becomes unstable, it was seen
that at the instability the chemical potential crosses the energies of the
normal modes at momenta associated with the signal and idler states.

At the effective chemical potential, the positive eigenvalue of the
distribution matrix of the polariton system diverges as $1/\omega$, and
so an effective temperature for the low energy modes was identified. The
effective temperature is proportional to the polariton pump mode occupation,
has global minima at ${\mathbf k}={\mathbf 0}$ and ${\mathbf k}=2{\mathbf
k}_p$, and a local maximum at ${\mathbf k}_p$. Remarkably, the system is
``condensing'' into modes characterised by a temperature near the lowest
possible. To relate to experimentally measurable quantities we computed the
incoherent luminescence, absorption and spectral weight for two pump strengths,
one just below the onset of instability and one where the ``normal state''
becomes stable again. We observe that close to the lower threshold the
signal (idler) state develops on a momentum ring rather than at a single
momentum as seen close to the upper threshold. The OPO state can therefore
be described simply in terms of only three dominant momentum modes if the
pump is decreased through the `upper threshold' of the instability.

In the OPO regime, calculating the incoherent luminescence around the three
modes shows that, although the spectra of small fluctuations are identical
around each mode, the Goldstone mode affects only the signal and idler states
and has little effect on the pump state. This is expected as the phase of the
pump state is fixed by the driving process and therefore only small phase
fluctuations are allowed. This is also in agreement with the observation
that the vortex- anti-vortex pairs across the BKT transition are present in
the signal and idler but not in the pump state. \cite{PhysRevX.5.041028}
It is clear that the occupations, as well as the forms of the excitation
spectra determine the properties of a state.

\acknowledgments{We would like to thank F. M. Marchetti and A. Berceanu for
helpful discussions. K.D. would like to thank J. M. Fellows for help with
early versions of the Fortran code used. We acknowledge support from EPSRC
(grants EP/I028900/2 and EP/K003623/2).}


\begin{thebibliography}{61}
\expandafter\ifx\csname natexlab\endcsname\relax\def\natexlab#1{#1}\fi
\expandafter\ifx\csname bibnamefont\endcsname\relax
  \def\bibnamefont#1{#1}\fi
\expandafter\ifx\csname bibfnamefont\endcsname\relax
  \def\bibfnamefont#1{#1}\fi
\expandafter\ifx\csname citenamefont\endcsname\relax
  \def\citenamefont#1{#1}\fi
\expandafter\ifx\csname url\endcsname\relax
  \def\url#1{\texttt{#1}}\fi
\expandafter\ifx\csname urlprefix\endcsname\relax\def\urlprefix{URL }\fi
\providecommand{\bibinfo}[2]{#2}
\providecommand{\eprint}[2][]{\url{#2}}

\bibitem[{\citenamefont{Carusotto and Ciuti}(2013)}]{RevModPhys.85.299}
\bibinfo{author}{\bibfnamefont{I.}~\bibnamefont{Carusotto}} \bibnamefont{and}
  \bibinfo{author}{\bibfnamefont{C.}~\bibnamefont{Ciuti}},
  \bibinfo{journal}{Rev. Mod. Phys.} \textbf{\bibinfo{volume}{85}},
  \bibinfo{pages}{299} (\bibinfo{year}{2013}).

\bibitem[{\citenamefont{Keeling and Berloff}(2011)}]{keeling}
\bibinfo{author}{\bibfnamefont{J.}~\bibnamefont{Keeling}} \bibnamefont{and}
  \bibinfo{author}{\bibfnamefont{N.~G.} \bibnamefont{Berloff}},
  \bibinfo{journal}{Contemporary Physics} \textbf{\bibinfo{volume}{52}},
  \bibinfo{pages}{131} (\bibinfo{year}{2011}).

\bibitem[{\citenamefont{Szyma\ifmmode~\acute{n}\else \'{n}\fi{}ska
  et~al.}(2006)\citenamefont{Szyma\ifmmode~\acute{n}\else \'{n}\fi{}ska,
  Keeling, and Littlewood}}]{PhysRevLett.96.230602}
\bibinfo{author}{\bibfnamefont{M.~H.} \bibnamefont{Szyma\ifmmode~\acute{n}\else
  \'{n}\fi{}ska}}, \bibinfo{author}{\bibfnamefont{J.}~\bibnamefont{Keeling}},
  \bibnamefont{and} \bibinfo{author}{\bibfnamefont{P.~B.}
  \bibnamefont{Littlewood}}, \bibinfo{journal}{Phys. Rev. Lett.}
  \textbf{\bibinfo{volume}{96}}, \bibinfo{pages}{230602}
  (\bibinfo{year}{2006}).

\bibitem[{\citenamefont{Szyma\ifmmode~\acute{n}\else \'{n}\fi{}ska
  et~al.}(2007)\citenamefont{Szyma\ifmmode~\acute{n}\else \'{n}\fi{}ska,
  Keeling, and Littlewood}}]{PhysRevB.75.195331}
\bibinfo{author}{\bibfnamefont{M.~H.} \bibnamefont{Szyma\ifmmode~\acute{n}\else
  \'{n}\fi{}ska}}, \bibinfo{author}{\bibfnamefont{J.}~\bibnamefont{Keeling}},
  \bibnamefont{and} \bibinfo{author}{\bibfnamefont{P.~B.}
  \bibnamefont{Littlewood}}, \bibinfo{journal}{Phys. Rev. B}
  \textbf{\bibinfo{volume}{75}}, \bibinfo{pages}{195331}
  (\bibinfo{year}{2007}).

\bibitem[{\citenamefont{Keeling et~al.}(2010)\citenamefont{Keeling,
  Szyma\'nska, and Littlewood}}]{NanoSciTech.146.Keldysh}
\bibinfo{author}{\bibfnamefont{J.}~\bibnamefont{Keeling}},
  \bibinfo{author}{\bibfnamefont{M.~H.} \bibnamefont{Szyma\'nska}},
  \bibnamefont{and} \bibinfo{author}{\bibfnamefont{P.~B.}
  \bibnamefont{Littlewood}}, in \emph{\bibinfo{booktitle}{Optical Generation
  and Control of Quantum Coherence in Semiconductor Nanostructures}}, edited by
  \bibinfo{editor}{\bibfnamefont{G.}~\bibnamefont{Slavcheva}} \bibnamefont{and}
  \bibinfo{editor}{\bibfnamefont{P.}~\bibnamefont{Roussignol}}
  (\bibinfo{publisher}{Springer Berlin Heidelberg}, \bibinfo{year}{2010}),
  Nanoscience and Technology, pp. \bibinfo{pages}{293--329}.

\bibitem[{\citenamefont{Proukakis et~al.}(2013)\citenamefont{Proukakis,
  Gardiner, Davis, and Szyma\'nska}}]{FINESSbook}
\bibinfo{editor}{\bibfnamefont{N.}~\bibnamefont{Proukakis}},
  \bibinfo{editor}{\bibfnamefont{S.}~\bibnamefont{Gardiner}},
  \bibinfo{editor}{\bibfnamefont{M.}~\bibnamefont{Davis}}, \bibnamefont{and}
  \bibinfo{editor}{\bibfnamefont{M.}~\bibnamefont{Szyma\'nska}}, eds.,
  \emph{\bibinfo{title}{Quantum Gases: Finite Temperature and Non-Equilibrium
  Dynamics}} (\bibinfo{publisher}{Imperial College Press},
  \bibinfo{year}{2013}).

\bibitem[{\citenamefont{Sieberer et~al.}(2013)\citenamefont{Sieberer, Huber,
  Altman, and Diehl}}]{PhysRevLett.110.195301}
\bibinfo{author}{\bibfnamefont{L.~M.} \bibnamefont{Sieberer}},
  \bibinfo{author}{\bibfnamefont{S.~D.} \bibnamefont{Huber}},
  \bibinfo{author}{\bibfnamefont{E.}~\bibnamefont{Altman}}, \bibnamefont{and}
  \bibinfo{author}{\bibfnamefont{S.}~\bibnamefont{Diehl}},
  \bibinfo{journal}{Phys. Rev. Lett.} \textbf{\bibinfo{volume}{110}},
  \bibinfo{pages}{195301} (\bibinfo{year}{2013}).

\bibitem[{\citenamefont{Buchhold et~al.}(2013)\citenamefont{Buchhold, Strack,
  Sachdev, and Diehl}}]{PhysRevA.87.063622}
\bibinfo{author}{\bibfnamefont{M.}~\bibnamefont{Buchhold}},
  \bibinfo{author}{\bibfnamefont{P.}~\bibnamefont{Strack}},
  \bibinfo{author}{\bibfnamefont{S.}~\bibnamefont{Sachdev}}, \bibnamefont{and}
  \bibinfo{author}{\bibfnamefont{S.}~\bibnamefont{Diehl}},
  \bibinfo{journal}{Phys. Rev. A} \textbf{\bibinfo{volume}{87}},
  \bibinfo{pages}{063622} (\bibinfo{year}{2013}).

\bibitem[{\citenamefont{Kasprzak et~al.}(2006)\citenamefont{Kasprzak, Richard,
  Kundermann, Baas, Jeambrun, Keeling, Marchetti, Szyma\'nska, Andr\'e, Staehli
  et~al.}}]{Nature443}
\bibinfo{author}{\bibfnamefont{J.}~\bibnamefont{Kasprzak}},
  \bibinfo{author}{\bibfnamefont{M.}~\bibnamefont{Richard}},
  \bibinfo{author}{\bibfnamefont{S.}~\bibnamefont{Kundermann}},
  \bibinfo{author}{\bibfnamefont{A.}~\bibnamefont{Baas}},
  \bibinfo{author}{\bibfnamefont{P.}~\bibnamefont{Jeambrun}},
  \bibinfo{author}{\bibfnamefont{J.~M.~J.} \bibnamefont{Keeling}},
  \bibinfo{author}{\bibfnamefont{F.~M.} \bibnamefont{Marchetti}},
  \bibinfo{author}{\bibfnamefont{M.~H.} \bibnamefont{Szyma\'nska}},
  \bibinfo{author}{\bibfnamefont{R.}~\bibnamefont{Andr\'e}},
  \bibinfo{author}{\bibfnamefont{J.~L.} \bibnamefont{Staehli}},
  \bibnamefont{et~al.}, \bibinfo{journal}{Nature}
  \textbf{\bibinfo{volume}{443}}, \bibinfo{pages}{409} (\bibinfo{year}{2006}).

\bibitem[{\citenamefont{Balili et~al.}(2007)\citenamefont{Balili, Hartwell,
  Snoke, Pfeiffer, and West}}]{Snoke3162007}
\bibinfo{author}{\bibfnamefont{R.}~\bibnamefont{Balili}},
  \bibinfo{author}{\bibfnamefont{V.}~\bibnamefont{Hartwell}},
  \bibinfo{author}{\bibfnamefont{D.~W.} \bibnamefont{Snoke}},
  \bibinfo{author}{\bibfnamefont{L.}~\bibnamefont{Pfeiffer}}, \bibnamefont{and}
  \bibinfo{author}{\bibfnamefont{K.}~\bibnamefont{West}},
  \bibinfo{journal}{Science} \textbf{\bibinfo{volume}{316}},
  \bibinfo{pages}{1007} (\bibinfo{year}{2007}).

\bibitem[{\citenamefont{Amo et~al.}(2009{\natexlab{a}})\citenamefont{Amo,
  Sanvitto, Laussy, Ballarini, del Valle, Martin, Lema\^itre, Bloch,
  Krizhanovskii, Skolnick et~al.}}]{Nature457}
\bibinfo{author}{\bibfnamefont{A.}~\bibnamefont{Amo}},
  \bibinfo{author}{\bibfnamefont{D.}~\bibnamefont{Sanvitto}},
  \bibinfo{author}{\bibfnamefont{F.~P.} \bibnamefont{Laussy}},
  \bibinfo{author}{\bibfnamefont{D.}~\bibnamefont{Ballarini}},
  \bibinfo{author}{\bibfnamefont{E.}~\bibnamefont{del Valle}},
  \bibinfo{author}{\bibfnamefont{M.~D.} \bibnamefont{Martin}},
  \bibinfo{author}{\bibfnamefont{A.}~\bibnamefont{Lema\^itre}},
  \bibinfo{author}{\bibfnamefont{J.}~\bibnamefont{Bloch}},
  \bibinfo{author}{\bibfnamefont{D.~N.} \bibnamefont{Krizhanovskii}},
  \bibinfo{author}{\bibfnamefont{M.~S.} \bibnamefont{Skolnick}},
  \bibnamefont{et~al.}, \bibinfo{journal}{Nature}
  \textbf{\bibinfo{volume}{457}}, \bibinfo{pages}{291}
  (\bibinfo{year}{2009}{\natexlab{a}}).

\bibitem[{\citenamefont{Amo et~al.}(2009{\natexlab{b}})\citenamefont{Amo,
  Lefr{\`e}re, Pigeon, Adrados, Ciuti, Carusotto, Houdr{\'e}, Giacobino, and
  Bramati}}]{amo2009superfluidity}
\bibinfo{author}{\bibfnamefont{A.}~\bibnamefont{Amo}},
  \bibinfo{author}{\bibfnamefont{J.}~\bibnamefont{Lefr{\`e}re}},
  \bibinfo{author}{\bibfnamefont{S.}~\bibnamefont{Pigeon}},
  \bibinfo{author}{\bibfnamefont{C.}~\bibnamefont{Adrados}},
  \bibinfo{author}{\bibfnamefont{C.}~\bibnamefont{Ciuti}},
  \bibinfo{author}{\bibfnamefont{I.}~\bibnamefont{Carusotto}},
  \bibinfo{author}{\bibfnamefont{R.}~\bibnamefont{Houdr{\'e}}},
  \bibinfo{author}{\bibfnamefont{E.}~\bibnamefont{Giacobino}},
  \bibnamefont{and} \bibinfo{author}{\bibfnamefont{A.}~\bibnamefont{Bramati}},
  \bibinfo{journal}{Nature Physics} \textbf{\bibinfo{volume}{5}},
  \bibinfo{pages}{805} (\bibinfo{year}{2009}{\natexlab{b}}).

\bibitem[{\citenamefont{Sanvitto et~al.}(2010)\citenamefont{Sanvitto,
  Marchetti, Szyma{\'n}ska, Tosi, Baudisch, Laussy, Krizhanovskii, Skolnick,
  Marrucci, Lema{\^\i}tre et~al.}}]{sanvitto2010persistent}
\bibinfo{author}{\bibfnamefont{D.}~\bibnamefont{Sanvitto}},
  \bibinfo{author}{\bibfnamefont{F.~M.} \bibnamefont{Marchetti}},
  \bibinfo{author}{\bibfnamefont{M.~H.} \bibnamefont{Szyma{\'n}ska}},
  \bibinfo{author}{\bibfnamefont{G.}~\bibnamefont{Tosi}},
  \bibinfo{author}{\bibfnamefont{M.}~\bibnamefont{Baudisch}},
  \bibinfo{author}{\bibfnamefont{F.~P.} \bibnamefont{Laussy}},
  \bibinfo{author}{\bibfnamefont{D.~N.} \bibnamefont{Krizhanovskii}},
  \bibinfo{author}{\bibfnamefont{M.~S.} \bibnamefont{Skolnick}},
  \bibinfo{author}{\bibfnamefont{L.}~\bibnamefont{Marrucci}},
  \bibinfo{author}{\bibfnamefont{A.}~\bibnamefont{Lema{\^\i}tre}},
  \bibnamefont{et~al.}, \bibinfo{journal}{Nature Physics}
  \textbf{\bibinfo{volume}{6}}, \bibinfo{pages}{527} (\bibinfo{year}{2010}).

\bibitem[{\citenamefont{Berceanu et~al.}(2015)\citenamefont{Berceanu, Dominici,
  Carusotto, Ballarini, Cancellieri, Gigli, Szyma\'{n}ska, Sanvitto, and
  Marchetti}}]{PhysRevB.92.035307}
\bibinfo{author}{\bibfnamefont{A.~C.} \bibnamefont{Berceanu}},
  \bibinfo{author}{\bibfnamefont{L.}~\bibnamefont{Dominici}},
  \bibinfo{author}{\bibfnamefont{I.}~\bibnamefont{Carusotto}},
  \bibinfo{author}{\bibfnamefont{D.}~\bibnamefont{Ballarini}},
  \bibinfo{author}{\bibfnamefont{E.}~\bibnamefont{Cancellieri}},
  \bibinfo{author}{\bibfnamefont{G.}~\bibnamefont{Gigli}},
  \bibinfo{author}{\bibfnamefont{M.~H.} \bibnamefont{Szyma\'{n}ska}},
  \bibinfo{author}{\bibfnamefont{D.}~\bibnamefont{Sanvitto}}, \bibnamefont{and}
  \bibinfo{author}{\bibfnamefont{F.~M.} \bibnamefont{Marchetti}},
  \bibinfo{journal}{Phys. Rev. B} \textbf{\bibinfo{volume}{92}},
  \bibinfo{pages}{035307} (\bibinfo{year}{2015}).

\bibitem[{\citenamefont{Nardin et~al.}(2011)\citenamefont{Nardin, Grosso,
  L{\'e}ger, Piȩtka, Morier-Genoud, and
  Deveaud-Pl{\'e}dran}}]{nardin2011hydrodynamic}
\bibinfo{author}{\bibfnamefont{G.}~\bibnamefont{Nardin}},
  \bibinfo{author}{\bibfnamefont{G.}~\bibnamefont{Grosso}},
  \bibinfo{author}{\bibfnamefont{Y.}~\bibnamefont{L{\'e}ger}},
  \bibinfo{author}{\bibfnamefont{B.}~\bibnamefont{Piȩtka}},
  \bibinfo{author}{\bibfnamefont{F.}~\bibnamefont{Morier-Genoud}},
  \bibnamefont{and}
  \bibinfo{author}{\bibfnamefont{B.}~\bibnamefont{Deveaud-Pl{\'e}dran}},
  \bibinfo{journal}{Nature Physics} \textbf{\bibinfo{volume}{7}},
  \bibinfo{pages}{635} (\bibinfo{year}{2011}).

\bibitem[{\citenamefont{Wertz et~al.}(2010)\citenamefont{Wertz, Ferrier,
  Solnyshkov, Johne, Sanvitto, Lema{\^\i}tre, Sagnes, Grousson, Kavokin,
  Senellart et~al.}}]{wertz2010spontaneous}
\bibinfo{author}{\bibfnamefont{E.}~\bibnamefont{Wertz}},
  \bibinfo{author}{\bibfnamefont{L.}~\bibnamefont{Ferrier}},
  \bibinfo{author}{\bibfnamefont{D.~D.} \bibnamefont{Solnyshkov}},
  \bibinfo{author}{\bibfnamefont{R.}~\bibnamefont{Johne}},
  \bibinfo{author}{\bibfnamefont{D.}~\bibnamefont{Sanvitto}},
  \bibinfo{author}{\bibfnamefont{A.}~\bibnamefont{Lema{\^\i}tre}},
  \bibinfo{author}{\bibfnamefont{I.}~\bibnamefont{Sagnes}},
  \bibinfo{author}{\bibfnamefont{R.}~\bibnamefont{Grousson}},
  \bibinfo{author}{\bibfnamefont{A.~V.} \bibnamefont{Kavokin}},
  \bibinfo{author}{\bibfnamefont{P.}~\bibnamefont{Senellart}},
  \bibnamefont{et~al.}, \bibinfo{journal}{Nature physics}
  \textbf{\bibinfo{volume}{6}}, \bibinfo{pages}{860} (\bibinfo{year}{2010}).

\bibitem[{\citenamefont{Sanvitto et~al.}(2011)\citenamefont{Sanvitto, Pigeon,
  Amo, Ballarini, De~Giorgi, Carusotto, Hivet, Pisanello, Sala, Guimaraes
  et~al.}}]{sanvitto2011all}
\bibinfo{author}{\bibfnamefont{D.}~\bibnamefont{Sanvitto}},
  \bibinfo{author}{\bibfnamefont{S.}~\bibnamefont{Pigeon}},
  \bibinfo{author}{\bibfnamefont{A.}~\bibnamefont{Amo}},
  \bibinfo{author}{\bibfnamefont{D.}~\bibnamefont{Ballarini}},
  \bibinfo{author}{\bibfnamefont{M.}~\bibnamefont{De~Giorgi}},
  \bibinfo{author}{\bibfnamefont{I.}~\bibnamefont{Carusotto}},
  \bibinfo{author}{\bibfnamefont{R.}~\bibnamefont{Hivet}},
  \bibinfo{author}{\bibfnamefont{F.}~\bibnamefont{Pisanello}},
  \bibinfo{author}{\bibfnamefont{V.~G.} \bibnamefont{Sala}},
  \bibinfo{author}{\bibfnamefont{P.~S.~S.} \bibnamefont{Guimaraes}},
  \bibnamefont{et~al.}, \bibinfo{journal}{Nature Photonics}
  \textbf{\bibinfo{volume}{5}}, \bibinfo{pages}{610} (\bibinfo{year}{2011}).

\bibitem[{\citenamefont{Grosso et~al.}(2011)\citenamefont{Grosso, Nardin,
  Morier-Genoud, L{\'e}ger, and Deveaud-Pl{\'e}dran}}]{grosso2011soliton}
\bibinfo{author}{\bibfnamefont{G.}~\bibnamefont{Grosso}},
  \bibinfo{author}{\bibfnamefont{G.}~\bibnamefont{Nardin}},
  \bibinfo{author}{\bibfnamefont{F.}~\bibnamefont{Morier-Genoud}},
  \bibinfo{author}{\bibfnamefont{Y.}~\bibnamefont{L{\'e}ger}},
  \bibnamefont{and}
  \bibinfo{author}{\bibfnamefont{B.}~\bibnamefont{Deveaud-Pl{\'e}dran}},
  \bibinfo{journal}{Physical review letters} \textbf{\bibinfo{volume}{107}},
  \bibinfo{pages}{245301} (\bibinfo{year}{2011}).

\bibitem[{\citenamefont{Amo et~al.}(2011)\citenamefont{Amo, Pigeon, Sanvitto,
  Sala, Hivet, Carusotto, Pisanello, Lem\'enager, Houdr\'e, Giacobino
  et~al.}}]{Amo3322011}
\bibinfo{author}{\bibfnamefont{A.}~\bibnamefont{Amo}},
  \bibinfo{author}{\bibfnamefont{S.}~\bibnamefont{Pigeon}},
  \bibinfo{author}{\bibfnamefont{D.}~\bibnamefont{Sanvitto}},
  \bibinfo{author}{\bibfnamefont{V.~G.} \bibnamefont{Sala}},
  \bibinfo{author}{\bibfnamefont{R.}~\bibnamefont{Hivet}},
  \bibinfo{author}{\bibfnamefont{I.}~\bibnamefont{Carusotto}},
  \bibinfo{author}{\bibfnamefont{F.}~\bibnamefont{Pisanello}},
  \bibinfo{author}{\bibfnamefont{G.}~\bibnamefont{Lem\'enager}},
  \bibinfo{author}{\bibfnamefont{R.}~\bibnamefont{Houdr\'e}},
  \bibinfo{author}{\bibfnamefont{E.}~\bibnamefont{Giacobino}},
  \bibnamefont{et~al.}, \bibinfo{journal}{Science}
  \textbf{\bibinfo{volume}{332}}, \bibinfo{pages}{1167} (\bibinfo{year}{2011}).

\bibitem[{\citenamefont{Hamp et~al.}(2015)\citenamefont{Hamp, Balin, Marchetti,
  Sanvitto, and Szyma{\'n}ska}}]{EPL.110.57006}
\bibinfo{author}{\bibfnamefont{J.~O.} \bibnamefont{Hamp}},
  \bibinfo{author}{\bibfnamefont{A.~K.} \bibnamefont{Balin}},
  \bibinfo{author}{\bibfnamefont{F.~M.} \bibnamefont{Marchetti}},
  \bibinfo{author}{\bibfnamefont{D.}~\bibnamefont{Sanvitto}}, \bibnamefont{and}
  \bibinfo{author}{\bibfnamefont{M.~H.} \bibnamefont{Szyma{\'n}ska}},
  \bibinfo{journal}{EPL (Europhysics Letters)} \textbf{\bibinfo{volume}{110}},
  \bibinfo{pages}{57006} (\bibinfo{year}{2015}).

\bibitem[{\citenamefont{Bobrovska and Matuszewski}(2015)}]{PhysRevB.92.035311}
\bibinfo{author}{\bibfnamefont{N.}~\bibnamefont{Bobrovska}} \bibnamefont{and}
  \bibinfo{author}{\bibfnamefont{M.}~\bibnamefont{Matuszewski}},
  \bibinfo{journal}{Phys. Rev. B} \textbf{\bibinfo{volume}{92}},
  \bibinfo{pages}{035311} (\bibinfo{year}{2015}).

\bibitem[{\citenamefont{Chiocchetta and Carusotto}(2013)}]{EPL.102.67007}
\bibinfo{author}{\bibfnamefont{A.}~\bibnamefont{Chiocchetta}} \bibnamefont{and}
  \bibinfo{author}{\bibfnamefont{I.}~\bibnamefont{Carusotto}},
  \bibinfo{journal}{EPL (Europhysics Letters)} \textbf{\bibinfo{volume}{102}},
  \bibinfo{pages}{67007} (\bibinfo{year}{2013}).

\bibitem[{\citenamefont{Savvidis et~al.}(2000)\citenamefont{Savvidis, Baumberg,
  Stevenson, Skolnick, Whittaker, and Roberts}}]{PhysRevLett.84.1547}
\bibinfo{author}{\bibfnamefont{P.~G.} \bibnamefont{Savvidis}},
  \bibinfo{author}{\bibfnamefont{J.~J.} \bibnamefont{Baumberg}},
  \bibinfo{author}{\bibfnamefont{R.~M.} \bibnamefont{Stevenson}},
  \bibinfo{author}{\bibfnamefont{M.~S.} \bibnamefont{Skolnick}},
  \bibinfo{author}{\bibfnamefont{D.~M.} \bibnamefont{Whittaker}},
  \bibnamefont{and} \bibinfo{author}{\bibfnamefont{J.~S.}
  \bibnamefont{Roberts}}, \bibinfo{journal}{Phys. Rev. Lett.}
  \textbf{\bibinfo{volume}{84}}, \bibinfo{pages}{1547} (\bibinfo{year}{2000}).

\bibitem[{\citenamefont{Stevenson et~al.}(2000)\citenamefont{Stevenson,
  Astratov, Skolnick, Whittaker, Emam-Ismail, Tartakovskii, Savvidis, Baumberg,
  and Roberts}}]{PhysRevLett.85.3680}
\bibinfo{author}{\bibfnamefont{R.~M.} \bibnamefont{Stevenson}},
  \bibinfo{author}{\bibfnamefont{V.~N.} \bibnamefont{Astratov}},
  \bibinfo{author}{\bibfnamefont{M.~S.} \bibnamefont{Skolnick}},
  \bibinfo{author}{\bibfnamefont{D.~M.} \bibnamefont{Whittaker}},
  \bibinfo{author}{\bibfnamefont{M.}~\bibnamefont{Emam-Ismail}},
  \bibinfo{author}{\bibfnamefont{A.~I.} \bibnamefont{Tartakovskii}},
  \bibinfo{author}{\bibfnamefont{P.~G.} \bibnamefont{Savvidis}},
  \bibinfo{author}{\bibfnamefont{J.~J.} \bibnamefont{Baumberg}},
  \bibnamefont{and} \bibinfo{author}{\bibfnamefont{J.~S.}
  \bibnamefont{Roberts}}, \bibinfo{journal}{Phys. Rev. Lett.}
  \textbf{\bibinfo{volume}{85}}, \bibinfo{pages}{3680} (\bibinfo{year}{2000}).

\bibitem[{\citenamefont{Baumberg et~al.}(2000)\citenamefont{Baumberg, Savvidis,
  Stevenson, Tartakovskii, Skolnick, Whittaker, and
  Roberts}}]{PhysRevB.62.16247}
\bibinfo{author}{\bibfnamefont{J.~J.} \bibnamefont{Baumberg}},
  \bibinfo{author}{\bibfnamefont{P.~G.} \bibnamefont{Savvidis}},
  \bibinfo{author}{\bibfnamefont{R.~M.} \bibnamefont{Stevenson}},
  \bibinfo{author}{\bibfnamefont{A.~I.} \bibnamefont{Tartakovskii}},
  \bibinfo{author}{\bibfnamefont{M.~S.} \bibnamefont{Skolnick}},
  \bibinfo{author}{\bibfnamefont{D.~M.} \bibnamefont{Whittaker}},
  \bibnamefont{and} \bibinfo{author}{\bibfnamefont{J.~S.}
  \bibnamefont{Roberts}}, \bibinfo{journal}{Phys. Rev. B}
  \textbf{\bibinfo{volume}{62}}, \bibinfo{pages}{R16247}
  (\bibinfo{year}{2000}).

\bibitem[{\citenamefont{Tartakovskii et~al.}(2002)\citenamefont{Tartakovskii,
  Krizhanovskii, Kurysh, Kulakovskii, Skolnick, and
  Roberts}}]{PhysRevB.65.081308}
\bibinfo{author}{\bibfnamefont{A.~I.} \bibnamefont{Tartakovskii}},
  \bibinfo{author}{\bibfnamefont{D.~N.} \bibnamefont{Krizhanovskii}},
  \bibinfo{author}{\bibfnamefont{D.~A.} \bibnamefont{Kurysh}},
  \bibinfo{author}{\bibfnamefont{V.~D.} \bibnamefont{Kulakovskii}},
  \bibinfo{author}{\bibfnamefont{M.~S.} \bibnamefont{Skolnick}},
  \bibnamefont{and} \bibinfo{author}{\bibfnamefont{J.~S.}
  \bibnamefont{Roberts}}, \bibinfo{journal}{Phys. Rev. B}
  \textbf{\bibinfo{volume}{65}}, \bibinfo{pages}{081308}
  (\bibinfo{year}{2002}).

\bibitem[{\citenamefont{Ciuti et~al.}(2001)\citenamefont{Ciuti, Schwendimann,
  and Quattropani}}]{PhysRevB.63.041303}
\bibinfo{author}{\bibfnamefont{C.}~\bibnamefont{Ciuti}},
  \bibinfo{author}{\bibfnamefont{P.}~\bibnamefont{Schwendimann}},
  \bibnamefont{and}
  \bibinfo{author}{\bibfnamefont{A.}~\bibnamefont{Quattropani}},
  \bibinfo{journal}{Phys. Rev. B} \textbf{\bibinfo{volume}{63}},
  \bibinfo{pages}{041303} (\bibinfo{year}{2001}).

\bibitem[{\citenamefont{Whittaker}(2001)}]{PhysRevB.63.193305}
\bibinfo{author}{\bibfnamefont{D.~M.} \bibnamefont{Whittaker}},
  \bibinfo{journal}{Phys. Rev. B} \textbf{\bibinfo{volume}{63}},
  \bibinfo{pages}{193305} (\bibinfo{year}{2001}).

\bibitem[{\citenamefont{Ciuti et~al.}(2003)\citenamefont{Ciuti, Schwendimann,
  and Quattropani}}]{SemicondSciTech.18.279}
\bibinfo{author}{\bibfnamefont{C.}~\bibnamefont{Ciuti}},
  \bibinfo{author}{\bibfnamefont{P.}~\bibnamefont{Schwendimann}},
  \bibnamefont{and}
  \bibinfo{author}{\bibfnamefont{A.}~\bibnamefont{Quattropani}},
  \bibinfo{journal}{Semicond. Sci. Technol.} \textbf{\bibinfo{volume}{18}},
  \bibinfo{pages}{S279} (\bibinfo{year}{2003}).

\bibitem[{\citenamefont{Carusotto and Ciuti}(2004)}]{PhysRevLett.93.166401}
\bibinfo{author}{\bibfnamefont{I.}~\bibnamefont{Carusotto}} \bibnamefont{and}
  \bibinfo{author}{\bibfnamefont{C.}~\bibnamefont{Ciuti}},
  \bibinfo{journal}{Phys. Rev. Lett.} \textbf{\bibinfo{volume}{93}},
  \bibinfo{pages}{166401} (\bibinfo{year}{2004}).

\bibitem[{\citenamefont{Whittaker}(2005)}]{PhysRevB.71.115301}
\bibinfo{author}{\bibfnamefont{D.~M.} \bibnamefont{Whittaker}},
  \bibinfo{journal}{Phys. Rev. B} \textbf{\bibinfo{volume}{71}},
  \bibinfo{pages}{115301} (\bibinfo{year}{2005}).

\bibitem[{\citenamefont{Ciuti and Carusotto}(2005)}]{PSSB:PSSB200560961}
\bibinfo{author}{\bibfnamefont{C.}~\bibnamefont{Ciuti}} \bibnamefont{and}
  \bibinfo{author}{\bibfnamefont{I.}~\bibnamefont{Carusotto}},
  \bibinfo{journal}{Phys. Status Solidi B} \textbf{\bibinfo{volume}{242}},
  \bibinfo{pages}{2224} (\bibinfo{year}{2005}), ISSN \bibinfo{issn}{1521-3951}.

\bibitem[{\citenamefont{Wouters and
  Carusotto}(2007{\natexlab{a}})}]{PhysRevB.75.075332}
\bibinfo{author}{\bibfnamefont{M.}~\bibnamefont{Wouters}} \bibnamefont{and}
  \bibinfo{author}{\bibfnamefont{I.}~\bibnamefont{Carusotto}},
  \bibinfo{journal}{Phys. Rev. B} \textbf{\bibinfo{volume}{75}},
  \bibinfo{pages}{075332} (\bibinfo{year}{2007}{\natexlab{a}}).

\bibitem[{\citenamefont{Wouters and
  Carusotto}(2007{\natexlab{b}})}]{PhysRevA.76.043807}
\bibinfo{author}{\bibfnamefont{M.}~\bibnamefont{Wouters}} \bibnamefont{and}
  \bibinfo{author}{\bibfnamefont{I.}~\bibnamefont{Carusotto}},
  \bibinfo{journal}{Phys. Rev. A} \textbf{\bibinfo{volume}{76}},
  \bibinfo{pages}{043807} (\bibinfo{year}{2007}{\natexlab{b}}).

\bibitem[{\citenamefont{Pitaevskii and
  Stringari}(2003)}]{BECPitaevskiiStringari}
\bibinfo{author}{\bibfnamefont{L.}~\bibnamefont{Pitaevskii}} \bibnamefont{and}
  \bibinfo{author}{\bibfnamefont{S.}~\bibnamefont{Stringari}},
  \emph{\bibinfo{title}{Bose-Einstein Condensation}}, no. \bibinfo{number}{116}
  in \bibinfo{series}{International Series of Monographs on Physics}
  (\bibinfo{publisher}{Oxford University Press}, \bibinfo{year}{2003}).

\bibitem[{\citenamefont{Dagvadorj et~al.}(2015)\citenamefont{Dagvadorj,
  Fellows, Matyja\ifmmode~\acute{s}\else \'{s}\fi{}kiewicz, Marchetti,
  Carusotto, and Szyma\ifmmode~\acute{n}\else
  \'{n}\fi{}ska}}]{PhysRevX.5.041028}
\bibinfo{author}{\bibfnamefont{G.}~\bibnamefont{Dagvadorj}},
  \bibinfo{author}{\bibfnamefont{J.~M.} \bibnamefont{Fellows}},
  \bibinfo{author}{\bibfnamefont{S.}~\bibnamefont{Matyja\ifmmode~\acute{s}\else
  \'{s}\fi{}kiewicz}}, \bibinfo{author}{\bibfnamefont{F.~M.}
  \bibnamefont{Marchetti}},
  \bibinfo{author}{\bibfnamefont{I.}~\bibnamefont{Carusotto}},
  \bibnamefont{and} \bibinfo{author}{\bibfnamefont{M.~H.}
  \bibnamefont{Szyma\ifmmode~\acute{n}\else \'{n}\fi{}ska}},
  \bibinfo{journal}{Phys. Rev. X} \textbf{\bibinfo{volume}{5}},
  \bibinfo{pages}{041028} (\bibinfo{year}{2015}).

\bibitem[{\citenamefont{Kamenev}(2005)}]{Kamenev}
\bibinfo{author}{\bibfnamefont{A.}~\bibnamefont{Kamenev}}, in
  \emph{\bibinfo{booktitle}{Les Houches, Session LXXXI, 2004 Nanophysics:
  Coherence and Transport}}, edited by
  \bibinfo{editor}{\bibfnamefont{H.}~\bibnamefont{Bouchiat}},
  \bibinfo{editor}{\bibfnamefont{Y.}~\bibnamefont{Gefen}},
  \bibinfo{editor}{\bibfnamefont{G.}~\bibnamefont{Montambaux}},
  \bibnamefont{and} \bibinfo{editor}{\bibfnamefont{J.}~\bibnamefont{Dalibard}}
  (\bibinfo{publisher}{Elsevier}, \bibinfo{year}{2005}).

\bibitem[{\citenamefont{Altland and Simons}(2010)}]{AltlandSimons}
\bibinfo{author}{\bibfnamefont{A.}~\bibnamefont{Altland}} \bibnamefont{and}
  \bibinfo{author}{\bibfnamefont{B.}~\bibnamefont{Simons}},
  \emph{\bibinfo{title}{Condensed Matter Field Theory}}
  (\bibinfo{publisher}{Cambridge University Press}, \bibinfo{year}{2010}),
  \bibinfo{edition}{2nd} ed.

\bibitem[{\citenamefont{Kamenev}(2011)}]{KamenevBook}
\bibinfo{author}{\bibfnamefont{A.}~\bibnamefont{Kamenev}},
  \emph{\bibinfo{title}{Field theory of non-equilibrium systems}}
  (\bibinfo{publisher}{Cambridge University Press}, \bibinfo{year}{2011}).

\bibitem[{\citenamefont{Torre et~al.}(2013)\citenamefont{Torre, Diehl, Lukin,
  Sachdev, and Strack}}]{PhysRevA.87.023831}
\bibinfo{author}{\bibfnamefont{E.~G.~D.} \bibnamefont{Torre}},
  \bibinfo{author}{\bibfnamefont{S.}~\bibnamefont{Diehl}},
  \bibinfo{author}{\bibfnamefont{M.~D.} \bibnamefont{Lukin}},
  \bibinfo{author}{\bibfnamefont{S.}~\bibnamefont{Sachdev}}, \bibnamefont{and}
  \bibinfo{author}{\bibfnamefont{P.}~\bibnamefont{Strack}},
  \bibinfo{journal}{Phys. Rev. A} \textbf{\bibinfo{volume}{87}},
  \bibinfo{pages}{023831} (\bibinfo{year}{2013}).

\bibitem[{\citenamefont{de~Leeuw et~al.}(2013)\citenamefont{de~Leeuw, Stoof,
  and Duine}}]{PhysRevA.88.033829}
\bibinfo{author}{\bibfnamefont{A.-W.} \bibnamefont{de~Leeuw}},
  \bibinfo{author}{\bibfnamefont{H.~T.~C.} \bibnamefont{Stoof}},
  \bibnamefont{and} \bibinfo{author}{\bibfnamefont{R.~A.} \bibnamefont{Duine}},
  \bibinfo{journal}{Phys. Rev. A} \textbf{\bibinfo{volume}{88}},
  \bibinfo{pages}{033829} (\bibinfo{year}{2013}).

\bibitem[{\citenamefont{Gopalakrishnan
  et~al.}(2010)\citenamefont{Gopalakrishnan, Lev, and
  Goldbart}}]{PhysRevA.82.043612}
\bibinfo{author}{\bibfnamefont{S.}~\bibnamefont{Gopalakrishnan}},
  \bibinfo{author}{\bibfnamefont{B.~L.} \bibnamefont{Lev}}, \bibnamefont{and}
  \bibinfo{author}{\bibfnamefont{P.~M.} \bibnamefont{Goldbart}},
  \bibinfo{journal}{Phys. Rev. A} \textbf{\bibinfo{volume}{82}},
  \bibinfo{pages}{043612} (\bibinfo{year}{2010}).

\bibitem[{\citenamefont{Yamaguchi et~al.}(2013)\citenamefont{Yamaguchi, Kamide,
  Nii, Ogawa, and Yamamoto}}]{PhysRevLett.111.026404}
\bibinfo{author}{\bibfnamefont{M.}~\bibnamefont{Yamaguchi}},
  \bibinfo{author}{\bibfnamefont{K.}~\bibnamefont{Kamide}},
  \bibinfo{author}{\bibfnamefont{R.}~\bibnamefont{Nii}},
  \bibinfo{author}{\bibfnamefont{T.}~\bibnamefont{Ogawa}}, \bibnamefont{and}
  \bibinfo{author}{\bibfnamefont{Y.}~\bibnamefont{Yamamoto}},
  \bibinfo{journal}{Phys. Rev. Lett.} \textbf{\bibinfo{volume}{111}},
  \bibinfo{pages}{026404} (\bibinfo{year}{2013}).

\bibitem[{\citenamefont{Yamaguchi et~al.}(2012)\citenamefont{Yamaguchi, Kamide,
  and Ogawa}}]{NewJPhys.14.065001}
\bibinfo{author}{\bibfnamefont{M.}~\bibnamefont{Yamaguchi}},
  \bibinfo{author}{\bibfnamefont{K.}~\bibnamefont{Kamide}}, \bibnamefont{and}
  \bibinfo{author}{\bibfnamefont{Y.}~\bibnamefont{Ogawa},
  \bibfnamefont{T.~Yamamoto}}, \bibinfo{journal}{New Journal of Physics}
  \textbf{\bibinfo{volume}{14}}, \bibinfo{pages}{065001}
  (\bibinfo{year}{2012}).

\bibitem[{\citenamefont{Kavokin et~al.}(2007)\citenamefont{Kavokin, Baumberg,
  Malpuech, and Laussy}}]{Microcavities}
\bibinfo{author}{\bibfnamefont{A.~V.} \bibnamefont{Kavokin}},
  \bibinfo{author}{\bibfnamefont{J.~J.} \bibnamefont{Baumberg}},
  \bibinfo{author}{\bibfnamefont{G.}~\bibnamefont{Malpuech}}, \bibnamefont{and}
  \bibinfo{author}{\bibfnamefont{F.~P.} \bibnamefont{Laussy}},
  \emph{\bibinfo{title}{Microcavities}} (\bibinfo{publisher}{Oxford University
  Press}, \bibinfo{year}{2007}).

\bibitem[{\citenamefont{Hopfield}(1958)}]{PhysRev.112.1555}
\bibinfo{author}{\bibfnamefont{J.~J.} \bibnamefont{Hopfield}},
  \bibinfo{journal}{Phys. Rev.} \textbf{\bibinfo{volume}{112}},
  \bibinfo{pages}{1555} (\bibinfo{year}{1958}).

\bibitem[{\citenamefont{Deng et~al.}(2010)\citenamefont{Deng, Haug, and
  Yamamoto}}]{RevModPhys.82.1489}
\bibinfo{author}{\bibfnamefont{H.}~\bibnamefont{Deng}},
  \bibinfo{author}{\bibfnamefont{H.}~\bibnamefont{Haug}}, \bibnamefont{and}
  \bibinfo{author}{\bibfnamefont{Y.}~\bibnamefont{Yamamoto}},
  \bibinfo{journal}{Rev. Mod. Phys.} \textbf{\bibinfo{volume}{82}},
  \bibinfo{pages}{1489} (\bibinfo{year}{2010}).

\bibitem[{\citenamefont{Marchetti and
  Szyma\'{n}ska}(2012)}]{FMMMHSVorticesinOPOChapt}
\bibinfo{author}{\bibfnamefont{F.~M.} \bibnamefont{Marchetti}}
  \bibnamefont{and} \bibinfo{author}{\bibfnamefont{M.~H.}
  \bibnamefont{Szyma\'{n}ska}}, in \emph{\bibinfo{booktitle}{Exciton Polaritons
  in Microcavities}}, edited by
  \bibinfo{editor}{\bibfnamefont{V.}~\bibnamefont{Timofeev}} \bibnamefont{and}
  \bibinfo{editor}{\bibfnamefont{D.}~\bibnamefont{Sanvitto}}
  (\bibinfo{publisher}{Springer Berlin Heidelberg}, \bibinfo{year}{2012}), vol.
  \bibinfo{volume}{172} of \emph{\bibinfo{series}{Springer Series in
  Solid-State Sciences}}, pp. \bibinfo{pages}{173--213}, ISBN
  \bibinfo{isbn}{978-3-642-24185-7}.

\bibitem[{\citenamefont{Butt\'e et~al.}(2003)\citenamefont{Butt\'e, Skolnick,
  Whittaker, Bajoni, and Roberts}}]{PhysRevB.68.115325}
\bibinfo{author}{\bibfnamefont{R.}~\bibnamefont{Butt\'e}},
  \bibinfo{author}{\bibfnamefont{M.~S.} \bibnamefont{Skolnick}},
  \bibinfo{author}{\bibfnamefont{D.~M.} \bibnamefont{Whittaker}},
  \bibinfo{author}{\bibfnamefont{D.}~\bibnamefont{Bajoni}}, \bibnamefont{and}
  \bibinfo{author}{\bibfnamefont{J.~S.} \bibnamefont{Roberts}},
  \bibinfo{journal}{Phys. Rev. B} \textbf{\bibinfo{volume}{68}},
  \bibinfo{pages}{115325} (\bibinfo{year}{2003}).

\bibitem[{\citenamefont{Landau and Lifshitz}(1965)}]{LandauV3}
\bibinfo{author}{\bibfnamefont{L.~D.} \bibnamefont{Landau}} \bibnamefont{and}
  \bibinfo{author}{\bibfnamefont{E.~M.} \bibnamefont{Lifshitz}},
  \emph{\bibinfo{title}{Quantum Mechanics}} (\bibinfo{publisher}{Pergamon
  Press}, \bibinfo{year}{1965}).

\bibitem[{\citenamefont{Carusotto and Ciuti}(2005)}]{PhysRevB.72.125335}
\bibinfo{author}{\bibfnamefont{I.}~\bibnamefont{Carusotto}} \bibnamefont{and}
  \bibinfo{author}{\bibfnamefont{C.}~\bibnamefont{Ciuti}},
  \bibinfo{journal}{Phys. Rev. B} \textbf{\bibinfo{volume}{72}},
  \bibinfo{pages}{125335} (\bibinfo{year}{2005}).

\bibitem[{\citenamefont{Marchetti et~al.}(2007)\citenamefont{Marchetti,
  Keeling, Szyma\ifmmode~\acute{n}\else \'{n}\fi{}ska, and
  Littlewood}}]{PhysRevB.76.115326}
\bibinfo{author}{\bibfnamefont{F.~M.} \bibnamefont{Marchetti}},
  \bibinfo{author}{\bibfnamefont{J.}~\bibnamefont{Keeling}},
  \bibinfo{author}{\bibfnamefont{M.~H.}
  \bibnamefont{Szyma\ifmmode~\acute{n}\else \'{n}\fi{}ska}}, \bibnamefont{and}
  \bibinfo{author}{\bibfnamefont{P.~B.} \bibnamefont{Littlewood}},
  \bibinfo{journal}{Phys. Rev. B} \textbf{\bibinfo{volume}{76}},
  \bibinfo{pages}{115326} (\bibinfo{year}{2007}).

\bibitem[{\citenamefont{Altman et~al.}(2015)\citenamefont{Altman, Sieberer,
  Chen, Diehl, and Toner}}]{PhysRevX.5.011017}
\bibinfo{author}{\bibfnamefont{E.}~\bibnamefont{Altman}},
  \bibinfo{author}{\bibfnamefont{L.~M.} \bibnamefont{Sieberer}},
  \bibinfo{author}{\bibfnamefont{L.}~\bibnamefont{Chen}},
  \bibinfo{author}{\bibfnamefont{S.}~\bibnamefont{Diehl}}, \bibnamefont{and}
  \bibinfo{author}{\bibfnamefont{J.}~\bibnamefont{Toner}},
  \bibinfo{journal}{Physical Review X} \textbf{\bibinfo{volume}{5}},
  \bibinfo{pages}{011017} (\bibinfo{year}{2015}).

\bibitem[{\citenamefont{Langbein}(2004)}]{PhysRevB.70.205301}
\bibinfo{author}{\bibfnamefont{W.}~\bibnamefont{Langbein}},
  \bibinfo{journal}{Phys. Rev. B} \textbf{\bibinfo{volume}{70}},
  \bibinfo{pages}{205301} (\bibinfo{year}{2004}).

\bibitem[{\citenamefont{Baas et~al.}(2004)\citenamefont{Baas, Karr, Eleuch, and
  Giacobino}}]{PhysRevA.69.023809}
\bibinfo{author}{\bibfnamefont{A.}~\bibnamefont{Baas}},
  \bibinfo{author}{\bibfnamefont{J.~P.} \bibnamefont{Karr}},
  \bibinfo{author}{\bibfnamefont{H.}~\bibnamefont{Eleuch}}, \bibnamefont{and}
  \bibinfo{author}{\bibfnamefont{E.}~\bibnamefont{Giacobino}},
  \bibinfo{journal}{Phys. Rev. A} \textbf{\bibinfo{volume}{69}},
  \bibinfo{pages}{023809} (\bibinfo{year}{2004}).

\bibitem[{\citenamefont{Song et~al.}(2005)\citenamefont{Song, Wang, and
  Makse}}]{Song15022005}
\bibinfo{author}{\bibfnamefont{C.}~\bibnamefont{Song}},
  \bibinfo{author}{\bibfnamefont{P.}~\bibnamefont{Wang}}, \bibnamefont{and}
  \bibinfo{author}{\bibfnamefont{H.~A.} \bibnamefont{Makse}},
  \bibinfo{journal}{Proceedings of the National Academy of Sciences of the
  United States of America} \textbf{\bibinfo{volume}{102}},
  \bibinfo{pages}{2299} (\bibinfo{year}{2005}).

\bibitem[{\citenamefont{Bouchbinder and Langer}(2009)}]{PhysRevE.80.031132}
\bibinfo{author}{\bibfnamefont{E.}~\bibnamefont{Bouchbinder}} \bibnamefont{and}
  \bibinfo{author}{\bibfnamefont{J.~S.} \bibnamefont{Langer}},
  \bibinfo{journal}{Phys. Rev. E} \textbf{\bibinfo{volume}{80}},
  \bibinfo{pages}{031132} (\bibinfo{year}{2009}).

\bibitem[{\citenamefont{Sollich and Fielding}(2002)}]{JPhys.CondMatt.14.1683}
\bibinfo{author}{\bibfnamefont{P.}~\bibnamefont{Sollich}} \bibnamefont{and}
  \bibinfo{author}{\bibfnamefont{P.}~\bibnamefont{Fielding},
  \bibfnamefont{S.and~Mayer}}, \bibinfo{journal}{Journal of Physics: Condensed
  Matter} \textbf{\bibinfo{volume}{14}}, \bibinfo{pages}{1683}
  (\bibinfo{year}{2002}).

\bibitem[{\citenamefont{Cugliandolo et~al.}(1997)\citenamefont{Cugliandolo,
  Kurchan, and Peliti}}]{PhysRevE.55.3898}
\bibinfo{author}{\bibfnamefont{L.~F.} \bibnamefont{Cugliandolo}},
  \bibinfo{author}{\bibfnamefont{J.}~\bibnamefont{Kurchan}}, \bibnamefont{and}
  \bibinfo{author}{\bibfnamefont{L.}~\bibnamefont{Peliti}},
  \bibinfo{journal}{Phys. Rev. E} \textbf{\bibinfo{volume}{55}},
  \bibinfo{pages}{3898} (\bibinfo{year}{1997}).

\bibitem[{\citenamefont{Sieberer et~al.}(2014)\citenamefont{Sieberer, Huber,
  Altman, and Diehl}}]{PhysRevB.89.134310}
\bibinfo{author}{\bibfnamefont{L.~M.} \bibnamefont{Sieberer}},
  \bibinfo{author}{\bibfnamefont{S.~D.} \bibnamefont{Huber}},
  \bibinfo{author}{\bibfnamefont{E.}~\bibnamefont{Altman}}, \bibnamefont{and}
  \bibinfo{author}{\bibfnamefont{S.}~\bibnamefont{Diehl}},
  \bibinfo{journal}{Phys. Rev. B} \textbf{\bibinfo{volume}{89}},
  \bibinfo{pages}{134310} (\bibinfo{year}{2014}).

\bibitem[{\citenamefont{Maghrebi and Gorshkov}(2015)}]{ArXiv:1507.01939}
\bibinfo{author}{\bibfnamefont{M.~F.} \bibnamefont{Maghrebi}} \bibnamefont{and}
  \bibinfo{author}{\bibfnamefont{A.~V.} \bibnamefont{Gorshkov}},
  \bibinfo{journal}{arXiv preprint arXiv:1507.01939} (\bibinfo{year}{2015}).

\end{thebibliography}
\end{document}